\documentclass[prb,aps,twocolumn,floatfix,amsmath,amssymb,tightenlines, showpacs, superscriptaddress]{revtex4}
\usepackage[pdftex]{graphicx}
 \usepackage{amsmath}
\usepackage[utf8x]{inputenc}
\usepackage{MnSymbol}
\usepackage[T1]{fontenc}
\usepackage{amssymb}
\usepackage{flushend}
 \usepackage{tikz}
\usetikzlibrary{decorations.markings}
\usepackage{amsfonts}
\usepackage{bm}
 \usepackage{amsmath} 
\usepackage[breaklinks=true,colorlinks=true,linkcolor=blue,urlcolor=blue,citecolor=blue]{hyperref}
\usepackage{amsfonts} 
\usepackage{amssymb, mathrsfs}
\usepackage{braket}
\usepackage{graphicx} 

\usepackage{bbm}
\def\beq{\begin{equation}}
\def\eeq{\end{equation}}
\def\bsp{\begin{split}}
\def\esp{\end{split}}
\def\bea{\begin{eqnarray}}
\def\eea{\end{eqnarray}}
\def\ba{\begin{array}}
\def\ea{\end{array}}

\def\dg{\dagger}

\def\lb{\left(}
\def\rb{\right)}

\def\l.{\left.}
\def\r.{\right.}

\def\ra{\rangle}
\def\la{\langle}

\def\bo{\bold{k}}

\begin{document}

\title{ Ground state properties of quantum  triangular ice}
\author{S. A. Owerre}
\email{sowerre@perimeterinstitute.ca}
\affiliation{Perimeter Institute for Theoretical Physics, 31 Caroline St. N., Waterloo, Ontario N2L 2Y5, Canada.}
\affiliation{African Institute for Mathematical Sciences, 6 Melrose Road, Muizenberg, Cape Town 7945, South Africa.}

%
%
%

\begin{abstract}
 Motivated by recent quantum Monte Carlo (QMC) simulations of the quantum Kagome ice model by Juan Carrasquilla, {\it et al.}, [Nature Communications, {\bf 6}, 7421, (2015)], we study  the ground state properties of this model on the triangular lattice. In the presence of a magnetic field $h$, the Hamiltonian possesses competing interactions between a $Z_2$-invariant easy-axis ferromagnetic interaction $J_{\pm\pm}$ and a frustrated Ising term  $J_z$. As in the U(1)-invariant model, we obtain four classical distinctive phases, however, the classical phases in the $Z_2$-invariant model are different. They are  as follows: a  fully polarized (FP) ferromagnet  for large $h$, an easy-axis canted ferromagnet (CFM) with broken $Z_2$ symmetry for small $h$ and dominant $J_{\pm\pm}$, a {\it ferrosolid}  phase with broken translational and $Z_2$ symmetries for small $h$ and dominant $J_{z}$, and two lobes with $m=\langle S_z\rangle=\pm 1/6$ for small $h$ and dominant $J_{z}$.  We show that quantum fluctuations are suppressed in this model, hence the large-$S$ expansion gives an accurate picture of the ground state properties.  When quantum fluctuations are introduced, we show that the {\it ferrosolid} state is the ground state in the dominant Ising limit at zero magnetic field. It remains robust for $J_z\to\infty$. With nonzero magnetic field the classical lobes acquire a finite magnetic susceptibility with no $S_z$-order. We present the trends of the ground state energy and the magnetizations. We also present a detail analysis of the CFM.  

\end{abstract}


\maketitle

\section {Introduction}   Quantum spin ice (QSI) on three-dimensional (3D) pyrochlore lattice is a subject of considerable interest in condensed matter physics.\cite{Huang, juan, gin, zhi,fen,kim, you, sun1, sun2,sun3, sun4, sun4a, sun5, sun6, sun7} In this quantum system, the spins are ordered in a two-in-two-out pattern reminiscent of hydrogen atoms in  water ice.  Competing interactions between a classically frustrated Ising coupling\cite{balent} and ferromagnetic quantum fluctuations give rise to fascinating  rich quantum phases.  Huang, Chen, and Hermele\cite{Huang} recently proposed an alternative simplified Hamiltonian that captures the physics of QSI on 3D pyrochlore lattice.

Recently, Carrasquilla {\it et~al.},\cite{juan}  showed that in the presence of a [111] crystallographic field, the 3D QSI model can be mapped onto a 2D frustrated system on the kagome lattice. Using a non-perturbative, unbiased QMC simulations, they uncovered the low-temperature quantum phase diagram on the kagome lattice.  The model studied by  Carrasquilla {\it et~al.}, \cite{juan} is termed  {\it quantum Kagome ice} and it is given explicitly by
 \begin{align}
&H= J_{\pm\pm}\sum_{\la lm\ra}\lb  -S_{l}^xS_{m}^x +S_{l}^yS_{m}^y\rb+ J_z\sum_{\la lm\ra} S_{l}^zS_{m}^z-h\sum_l S_l^z,
\label{k1}
\end{align}
The sign of $J_{\pm\pm}$ in Eq.~\eqref{k1} can be positive or negative under a $\pi/2$-rotation about the $z$-axis:  $S_{l,m}^{x} \to -S_{l,m}^{y},~S_{l,m}^{y,z} \to S_{l,m}^{x,z}$. The Hamiltonian, Eq.~\eqref{k1},  exhibits only $Z_2$ symmetry in the $x$-$y$ plane by a $\pi$-rotation about the $z$-axis: $S_{l,m}^{x,y} \to -S_{l,m}^{x,y}$; $S_{l,m}^z\to S_{l,m}^z$. The $Z_2$ symmetry of Eq.~\eqref{k1}  is synonymous with the fact that  the total $S_z$ is not conserved. For bipartite lattices, Eq.~\eqref{k1} can be transformed to a U(1)-invariant XXZ model \cite{fa, fa1,has0,has,has1, has2,kle1,isa, isa1,isa2,isa3, roger, xu,dar} by a $\pi$-rotation about the $x$-axis on one sublattice, $S_m^x\to S_m^x$,  $S_m^{y,z}\to -S_m^{y,z}$. Thus,  the effects of the $Z_2$ symmetry  of Eq.~\eqref{k1} is well-pronounced only on non-bipartite lattices where such rotation is invalid. In other words,  Eq.~\eqref{k1} cannot be mapped to a U(1)-invariant XXZ model on non-bipartite lattice.

 One of the distinguishing features of  Eq.~\eqref{k1} is that the ferromagnetic coupling is  easy-axis (say, along the $x$-axis); whereas in  the U(1)-invariant XXZ model, \cite{fa, fa1,has0,has,has1, has2,kle1,isa, isa1,isa2,isa3, roger, xu,dar} the ferromagnetic coupling is  easy-plane, that is in the $x$-$y$ plane. The phase diagram of Eq.~\eqref{k1} on the kagome lattice\cite{juan} has been shown to be different from the phase diagram of the U(1)-invariant XXZ model (hard-core bosons) on the kagome lattice.\cite{isak} The three phases  obtained from Eq.~\eqref{k1}  on the kagome lattice are as follows:\cite{juan} a trivial fully polarized (FP) ferromagnet (FM)  which appears for large magnetic field;  a predominant $x$-easy-axis ferromagnet  at $h/J_z= 0$, which becomes canted for small $h/J_z\neq 0$ and  $J_{\pm\pm}/J_z>0.5$. Finally, a  disordered magnetized lobe with $m=\la S_z\ra \approx 1/6$, which promotes a putative quantum spin liquid (QSL) phase  and appear for nonzero small $h/J_z\neq 0$ and large Ising coupling, $J_{\pm\pm}/J_z<0.5$. The phase diagram is symmetric with respect to the sign of $h/J_z$.  We have recently complemented  the QMC results on the kagome lattice using an explicit analytical spin wave theory. \cite{sow1} We showed that spin wave theory captures quantitatively all the trends observed in QMC including the   disordered magnetized lobe. 
  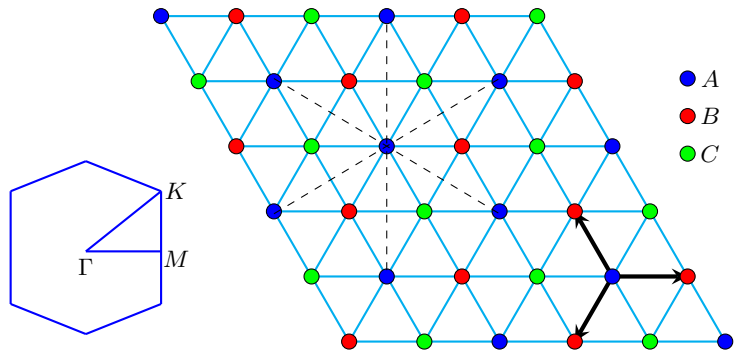
\begin{figure*}[ht]
\centering
\begin{tikzpicture}
\draw[thick,blue] (6.5,.5)--(6.5,2);
\draw[thick, blue] (6.5,2)--(5.5,2.4);
\draw[thick, blue] (5.5,2.4)--(4.5,2);
\draw[thick, blue] (4.5,2)--(4.5,.5);
\draw[thick, blue] (4.5,.5)--(5.5,.1);
\draw[thick, blue] (6.5,.5)--(5.5,.1);
\draw[thick, blue] (5.5,1.2)--(6.5,1.2);
\draw[thick, blue] (5.5,1.2)--(6.5,2);
\draw (5.5,1) node[]{$\Gamma$};
\draw (6.7,1.1) node[]{$M$};
\draw (6.7,2) node[]{$K$};
\draw[thick,cyan] (9,0)--(10,0);
\draw[thick,cyan] (9,0)--(9.5,0.866);
\draw[thick,cyan] (9,0)--(8.5,0.866);
\draw[thick,cyan] (10,0)--(11,0);
\draw[thick,cyan] (10,0)--(10.5,0.866);
\draw[thick,cyan] (10,0)--(9.5,0.866);
\draw[thick,cyan] (11,0)--(12,0);
\draw[thick,cyan] (11,0)--(11.5,0.866);
\draw[thick,cyan] (11,0)--(10.5,0.866);
\draw[thick,cyan] (12,0)--(13,0);
\draw[thick,cyan] (12,0)--(12.5,0.866);
\draw[->,>=stealth,ultra thick,
    ]
    (12.5,0.866)--(12,0);
\draw[thick,cyan] (12,0)--(11.5,0.866);
\draw[thick,cyan] (13,0)--(14,0);
\draw[thick,cyan] (13,0)--(13.5,0.866);
\draw[thick,cyan] (13,0)--(12.5,0.866);
\draw[thick,cyan] (14,0)--(13.5,0.866);
\draw[thick,cyan] (8.5,0.866)--(9.5, 0.866);
\draw[thick,cyan] (8.5,0.866)--(9,1.732);
\draw[thick,cyan] (8.5,0.866)--(8,1.732);
\draw[thick,cyan] (9.5,0.866)--(10.5, 0.866);
\draw[thick,cyan] (9.5,0.866)--(10,1.732);
\draw[thick,cyan] (9.5,0.866)--(9,1.732);
\draw[thick,cyan] (10.5,0.866)--(11.5, 0.866);
\draw[thick,cyan] (10.5,0.866)--(11,1.732);
\draw[thick,cyan] (10.5,0.866)--(10,1.732);
\draw[thick,cyan] (11.5,0.866)--(12.5, 0.866);
\draw[thick,cyan] (11.5,0.866)--(12,1.732);
\draw[thick,cyan] (11.5,0.866)--(11,1.732);
\draw[thick,cyan] (12.5,0.866)--(13.5, 0.866);
   \draw[->,>=stealth,ultra thick,
    ]
    (12.5,0.866)--(13.5, 0.866);
\draw[thick,cyan] (12.5,0.866)--(13,1.732);
\draw[thick,cyan] (12.5,0.866)--(12,1.732);
\draw[->,>=stealth,ultra thick,
    ]
    (12.5,0.866)--(12,1.732);
   
\draw[thick,cyan] (13.5,0.866)--(13,1.732);
\draw[thick,cyan] (8,1.732)--(9,1.732);
\draw[thick,cyan] (8,1.732)--(8.5,2.598);
\draw[thick,cyan] (8,1.732)--(7.5,2.598);
\draw[thick,cyan] (9,1.732)--(10,1.732);
\draw[thick,cyan] (9,1.732)--(9.5,2.598);
\draw[thick,cyan] (9,1.732)--(8.5,2.598);
\draw[thick,cyan] (10,1.732)--(11,1.732);
\draw[thick,cyan] (10,1.732)--(10.5,2.598);
\draw[thick,cyan] (10,1.732)--(9.5,2.598);
\draw[thick,cyan] (11,1.732)--(12,1.732);
\draw[thick,cyan] (11,1.732)--(11.5,2.598);
\draw[thick,cyan] (11,1.732)--(10.5,2.598);
\draw[thick,cyan] (12,1.732)--(13,1.732);
\draw[thick,cyan] (12,1.732)--(12.5,2.598);
\draw[thick,cyan] (12,1.732)--(11.5,2.598);
\draw[thick,cyan] (13,1.732)--(12.5,2.598);
\draw[thick,cyan] (7.5,2.598)--(8.5,2.598);
\draw[thick,cyan ] (7.5,2.598)--(8,3.464);
\draw[thick,cyan] (7.5,2.598)--(7,3.464);
\draw[thick,cyan] (8.5,2.598)--(9.5,2.598);
\draw[thick,cyan] (8.5,2.598)--(9,3.464);
\draw[thick,cyan] (8.5,2.598)--(8,3.464);
\draw[thick,cyan] (9.5,2.598)--(10.5,2.598);
\draw[thick,cyan] (9.5,2.598)--(10,3.464);
\draw[thick,cyan] (9.5,2.598)--(9,3.464);
\draw[thick,cyan] (10.5,2.598)--(11.5,2.598);
\draw[thick,cyan] (10.5,2.598)--(11,3.464);
\draw[thick,cyan] (10.5,2.598)--(10,3.464);
\draw[thick,cyan] (11.5,2.598)--(12.5,2.598);
\draw[thick,cyan] (11.5,2.598)--(12,3.464);
\draw[thick,cyan] (11.5,2.598)--(11,3.464);
\draw[thick,cyan] (12.5,2.598)--(12,3.464);
\draw[thick,cyan] (7,3.464)--(8,3.464);
\draw[thick,cyan] (7,3.464)--(7.5,4.330);
\draw[thick,cyan] (7,3.464)--(6.5,4.330);
\draw[thick,cyan] (8,3.464)--(9,3.464);
\draw[thick,cyan] (8,3.464)--(8.5,4.330);
\draw[thick,cyan] (8,3.464)--(7.5,4.330);
\draw[thick,cyan] (9,3.464)--(10,3.464);
\draw[thick,cyan] (9,3.464)--(9.5,4.330);
\draw[thick,cyan] (9,3.464)--(8.5,4.330);
\draw[thick,cyan] (10,3.464)--(11,3.464);
\draw[thick,cyan] (10,3.464)--(10.5,4.330);
\draw[thick,cyan] (10,3.464)--(9.5,4.330);
\draw[thick,cyan] (11,3.464)--(12,3.464);
\draw[thick,cyan] (11,3.464)--(11.5,4.330);
\draw[thick,cyan] (11,3.464)--(10.5,4.330);
\draw[thick,cyan] (12,3.464)--(11.5,4.330);
\draw[thick,cyan] (6.5,4.330)--(7.5,4.330);
\draw[thick,cyan] (7.5,4.330)--(8.5,4.330);
\draw[thick,cyan] (8.5,4.330)--(9.5,4.330);
\draw[thick,cyan] (9.5,4.330)--(10.5,4.330);
\draw[thick,cyan] (10.5,4.330)--(11.5,4.330);
\draw[fill= red] (9,0) circle (1mm);
\draw[fill= green] (10,0) circle (1mm);
\draw[fill= blue] (11,0) circle (1mm);
\draw[fill= red] (12,0) circle (1mm);
\draw[fill= green] (13,0) circle (1mm);
\draw[fill= blue] (14,0) circle (1mm);
\draw[fill= blue] (8,1.732) circle (1mm);
\draw[fill= red] (9,1.732) circle (1mm);
\draw[fill= green] (10,1.732) circle (1mm);
\draw[fill= blue] (11,1.732) circle (1mm);
\draw[fill= red] (12,1.732) circle (1mm);
\draw[fill= green] (13,1.732) circle (1mm);
\draw[fill= red] (7.5,2.598) circle (1mm);
\draw[fill= green] (8.5,2.598) circle (1mm);
\draw[fill= blue] (9.5,2.598) circle (1mm);
\draw[fill= red] (10.5,2.598) circle (1mm);
\draw[fill= green] (11.5,2.598) circle (1mm);
\draw[fill= blue] (12.5,2.598) circle (1mm);
\draw[fill= green] (7,3.464) circle (1mm);
\draw[fill= blue] (8,3.464) circle (1mm);
\draw[fill= red] (9,3.464) circle (1mm);
\draw[fill= green] (10,3.464) circle (1mm);
\draw[fill= blue] (11,3.464) circle (1mm);
\draw[fill= red] (12,3.464) circle (1mm);
\draw[fill= blue] (6.5,4.330) circle (1mm);
\draw[fill= red] (7.5,4.330) circle (1mm);
\draw[fill= green] (8.5,4.330) circle (1mm);
\draw[fill= blue] (9.5,4.330) circle (1mm);
\draw[fill= red] (10.5,4.330) circle (1mm);
\draw[fill= green] (11.5,4.330) circle (1mm);
\draw[fill= green] (8.5,0.866) circle (1mm);
\draw[fill= blue] (9.5,0.866) circle (1mm);
\draw[fill= red] (10.5,0.866) circle (1mm);
\draw[fill= green] (11.5,0.866) circle (1mm);
\draw[fill= blue] (12.5,0.866) circle (1mm);
\draw[fill= red] (13.5,0.866) circle (1mm);
\draw[dashed,black] (9.5,.95)--(9.5,4.3);
\draw[dashed, black] (8,1.7)--(11,3.5);
\draw[dashed, black] (8,3.5)--(11,1.7);
\draw[fill= blue] (13.5,3.5) circle (1mm);
\draw (13.8,3.5) node[]{$A$};
\draw[fill= red] (13.5,3) circle (1mm);
\draw (13.8,3) node[]{$B$};
\draw[fill= green] (13.5,2.5) circle (1mm);
\draw (13.8,2.5) node[]{$C$};
\end{tikzpicture}
\caption{(Color online) Triangular lattice with three-sublattice structure (right).  The corresponding one-sublattice Brillouin zone (left), with $\Gamma=(0,0)$; $M=(2\pi/3,0)$; $K=(2\pi/3,2\pi/3\sqrt{3})$. The nearest neighbour vectors (thick arrows) are $\bold{e}_1=(-\frac{1}{2}, -\frac{\sqrt 3}{2})$; $\bold{e}_2=(-\frac{1}{2}, \frac{\sqrt 3}{2})$; $\bold{e}_3=(1,0)$, and the next-nearest neighbour vectors (dashed line) are $\boldsymbol{\delta}_1=(\frac{3}{2}, \frac{\sqrt 3}{2})$; $\boldsymbol{\delta}_2=(\frac{3}{2}, -\frac{\sqrt 3}{2})$; $\boldsymbol{\delta}_3=(0,\sqrt{3})$.}\label{trif}
\end{figure*}

 In this paper, we consider  Eq.~\eqref{k1} on the triangular lattice. At the moment,  there is no existing QMC simulation of Eq.~\eqref{k1}  on the triangular lattice and  the nature of the unconventional phases has not been studied on the triangular lattice  both numerically and analytically. Thus, the analyses we present in this paper are not complementary to any existing numerical results. This paper, however,  serves as a guide to upcoming numerical simulations.  Spin wave theory is known to be suitable on the triangular lattice.  Also, the phase diagram of  the U(1)-invariant  XXZ model (hard-core boson) on the kagome lattice \cite{isak} is known to be different from that of the triangular lattice. \cite{fa, fa1,has0,has,has1, has2,kle1,isa, isa1,isa2,isa3, roger, xu,dar}  Hence, it is expedient to study the nature of the quantum phases of Eq.~\eqref{k1} on the triangular lattice and show their distinguishing features  from the well-studied U(1)-invariant XXZ model (hard-core boson).\cite{fa, fa1,has0,has,has1, has2,kle1,isa, isa1,isa2,isa3, roger, xu,dar} The motivation for using spin wave theory is that quantum fluctuations are suppressed in Eq.~\eqref{k1}, the occupation number $\la n_l\ra=\la b_l^\dagger b_l\ra$ characterizing the strength of quantum fluctuations appears to be very small. This is a consequence of the gapped nature of Eq.~\eqref{k1} as we will show later.  Hence,  spin wave theory works very well in the description of the ground state quantum properties of Eq.~\eqref{k1} on the triangular lattice.

We uncover four phases on the triangular lattice. In addition to the two ferromagnetic phases on the kagome lattice,  we uncover another phase selected via order-by-disorder mechanism in the frustrated regime,  $J_z\gg J_{\pm\pm}$ with $h=0$.  This phase is very robust even for $J_z\to\infty$ limit and it is called a {\it ferrosolid} state. It differs from the conventional  {\it supersolid} state,\cite{fa, fa1,has0,has,has1, has2,kle1,isa, isa1,isa2,isa3, roger, xu,dar}  as it simultaneously breaks two discrete symmetries --- translational and $Z_2$ symmetries. However, it retains the usual three-sublattice ferrimagnetic ordering pattern,\cite{has} $\la S_z\ra=(x,x,-x^\prime)$, with $x\neq x^\prime$, and the one-sublattice wave vector remains at $\bo=(\pm 4\pi/3,0)$. We also uncover a similar   magnetized lobe with $m=\la S_z\ra \approx 1/6$, which appear for small $h/J_z\neq 0$ and $J_{\pm\pm}/J_z<0.5$ as in the kagome lattice.\cite{juan,sow1}

This paper is organized as follows. In Sec.~\eqref{sec1} we present the classical phase diagram of Eq.~\eqref{k1} on the triangular lattice. We will show how the phase diagram differs  from its U(1) counterpart based on the broken symmetries of the Hamiltonian. In Sec.~\eqref{sec3} we introduce quantum fluctuations and perform the large-$S$ expansion of the Hamiltonian in Eq.~\eqref{k1} around each classical ground state. In Sec.~\eqref{sec3a} we present the evidence of suppression of quantum fluctuations through the selection of ground state at zero magnetic field. The energy spectra in each of the classical phases are presented in Sec.~\eqref{sec5}.  The effects of quantum fluctuations on the classical phase diagram and the order parameters at zero   and nonzero magnetic field are discussed in Sec.~\eqref{sec7}. We explore the easy-axis ferromagnetic phase in Sec.~\eqref{cfm}. Finally, we give a concluding remark in Sec.~\eqref{sec9}.

\section{Mean field Approximation}
\label{sec1}
In this section, we present the classical phase diagram of Eq.~\eqref{k1} on the triangular lattice.   In the mean field approximation or the large-$S$ limit, the spins can be approximated as classical vectors parameterized by a unit vector: $\bold{S}_l= S\hat{\bold{m}}_l$, where $\hat{\bold{m}}_l=\lb\sin\theta_l\cos\phi_l, \sin\theta_l\sin\phi_l,\cos\theta_l \rb$. 
 Adopting a three-sublattice configuration on the triangular lattice as depicted in Fig.~\eqref{trif}, the classical energy  is given by
\begin{align}
e_c &= \sum_{\alpha\beta}z_{\alpha\beta}\lb J_z \cos\theta_\alpha\cos\theta_{\beta}  -J_{\pm\pm}\sin\theta_\alpha\sin\theta_{\beta}\cos\phi_{\alpha\beta}\rb\label{cla}\\&\nonumber-h\sum_\alpha\cos\theta_\alpha, 
\end{align}
where we have taken the magnetic field to be of order $S$, hence $e_c=\mathcal{E}_c/NS^2$,  $\phi_{\alpha\beta}=\phi_\alpha+\phi_\beta$, and  $z_{\alpha\beta}$ is the number of $\alpha$ sublattice nearest to $\beta$ sublattice, and $\alpha,\beta=A,B,C$ are the sublattice indices. Here $N$ is the number of unit cells and $\mathcal{N}=3N$ is the total number of sites.  The $Z_2$-invariant classical energy differs  from the  U(1) counterpart by the azimuthal angles $\phi_{\alpha\beta}$.  In the convention we have chosen in Eq.~\eqref{cla}, ($J_{\pm\pm}>0$ and $J_z>0$) we consider $\phi_{\alpha\beta}=0$ since the mean field energy is coplanar.  For $h=0$, the regime $J_z\gg J_{\pm\pm}$ is frustrated. The classical ground states of Eq.~\eqref{k1} in this regime is degenerate, which exhibit a one-parameter  parameterization by $\theta_A$ with  $\theta_B=\delta-\epsilon$ and $\theta_C=\delta+\epsilon$: \cite{kle,kle1, pen, kle0} 
\begin{align}
 &\cos \delta= \frac{-\kappa\cos\theta_A}{\sqrt{1-\lb 1-\kappa^2\rb\cos^2\theta_A}},\\&\cos \epsilon= -\frac{\cos \delta}{(1-\kappa)\cos\theta_A}, \quad \kappa\neq 1,
 \label{onep}
 \end{align}
where  $\kappa=J_{\pm\pm}/J_z$.

 Application of a  nonzero magnetic field is known to lift the degenerate ground states, hence there exists a unique classical ground state for $h\neq0$.
\begin{figure}[ht]
\centering
\includegraphics[width=3in]{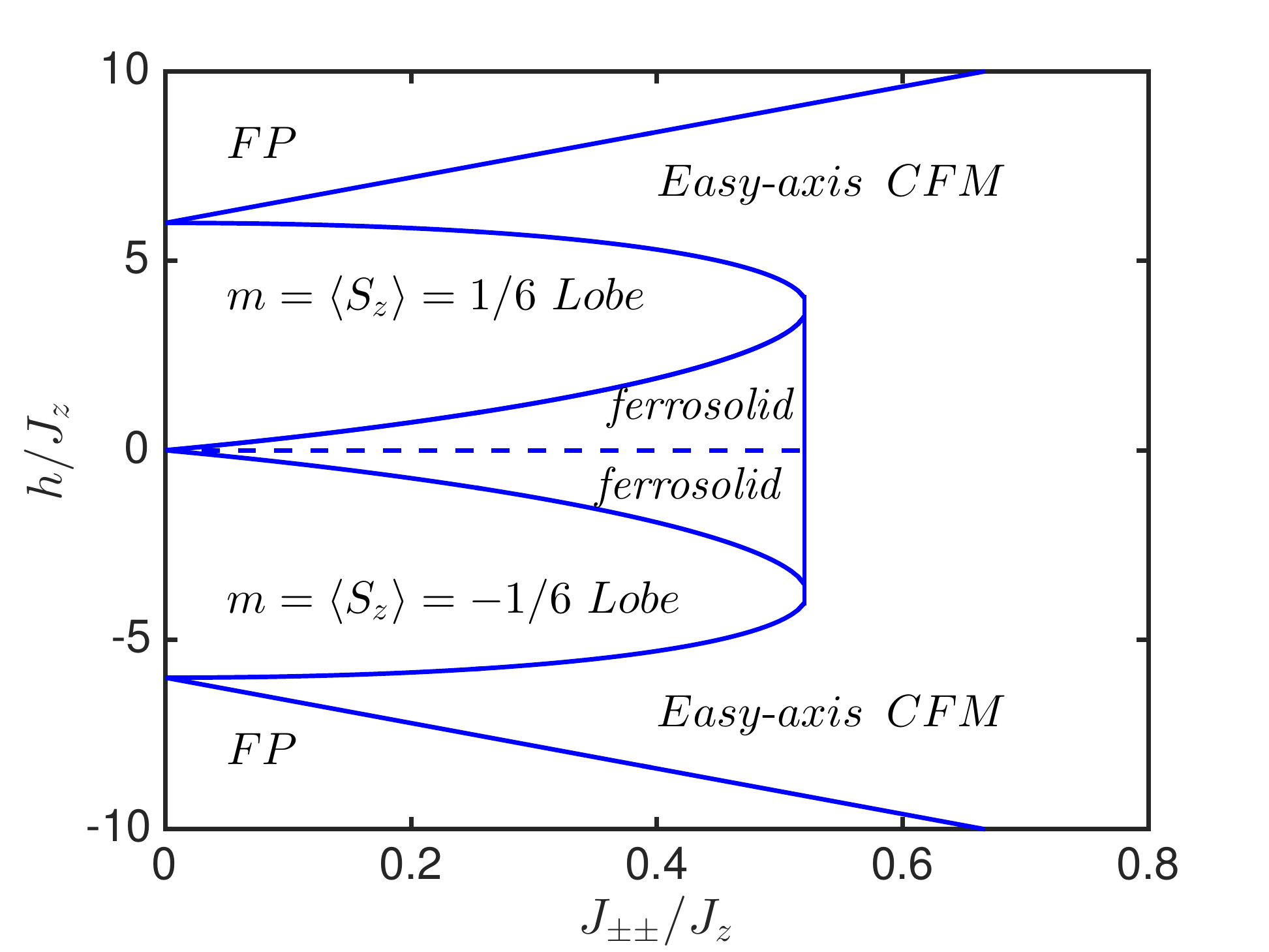}
\caption{The mean field phase diagram of  the $Z_2$-invariant model in Eq.~\eqref{k1}  as a function of $h/{J}_z$ and $J_{\pm\pm}/{J}_z$.}
\label{phase_lobe}
\end{figure}
The classical phase diagram is obtained by minimizing  Eq.~\eqref{cla} and considering the broken symmetries. From Fig.~\eqref{phase_lobe} we see that the classical phase diagram  has the same structure as   the U(1)-invariant model, \cite{fa, fa1,has0,has,has1, has2,kle1,isa, isa1,isa2,isa3, roger, xu,dar} but the interpretation of the phases are different.  The conventional superfluid phase is replaced with an easy-axis ferromagnetic ordered phase resulting from the spontaneously  broken $Z_2$ symmetry along the $x$-axis. This phase  is predominant in the limit $h=0$, $J_{\pm\pm}/ J_z>0.5$. As the magnetic field is increased from zero, the easy-axis ferromagnetic ordered phase becomes a canted ferromagnet (CFM). For large $\pm h/J_z$, there are two FP  ferromagnets along the $S_z$ direction.  The transition between the CFM and the FP occurs at $h=h_F= 6[J_{\pm\pm}+J_z]$. 
 
For  $J_{\pm\pm}/J_z<0.5$ and $h/J_z\neq 0$, there are two $m=\la S_z\ra=\pm 1/6$ lobes for positive and negative magnetic fields respectively. They have  a three-sublattice up-up-down and down-down-up patterns. As  Eq.~\eqref{k1}  is devoid of U(1) symmetry, the conventional {\it supersolid} (SS) phase\cite{fa, fa1,has0,has,has1, has2,kle1,isa, isa1,isa2,isa3, roger, xu,dar} is replaced with a {\it ferrosolid}  (FS) phase with broken translational and $Z_2$ symmetries. The {\it ferrosolid} state retains the usual three-sublattice ferrimagnetic ordering pattern,\cite{has} $\la S_z\ra=(x,x,-x^\prime)$, with $x\neq x^\prime$. The transition between the $ m=\la S_z\ra= 1/6$ and {\it ferrosolid}  occurs at $h=h_N=3\lb 2J_z +J_{\pm\pm} -{J}\rb/2$, with ${J}=\sqrt{4J_z(J_z-J_{\pm\pm})-7J_{\pm\pm}^2}$.

The classical phase diagram  in the U(1)-invariant model\cite{kle1}  generally have the same structure as its quantum phase diagram.\cite{fa, fa1,has0,has,has1, has2,isa, isa1,isa2,isa3, roger, xu,dar} The only difference is that the quantum phases appear in a smaller boundary of the parameter regime due to strong  quantum fluctuations. As we will show in the subsequent sections, quantum fluctuations are small in the $Z_2$-invariant model. Thus, the classical phase boundaries will be {\it approximately} the same as the quantum ones, since quantum fluctuations are small. In other words, quantum fluctuations will modify the boundaries very slightly.  This modification will be very small compared to the U(1) model as we have confirmed on the kagome lattice.\cite{sow1,juan}  Our main goals in this paper are  the modification of the  classical lobes  by the $Z_2$-invariant quantum  fluctuations and the trends of the ground state energy and the magnetizations at zero magnetic field. These are the main properties investigated on the kagome lattice.\cite{sow1,juan}

 \section{Spin wave theory}
\label{sec3}
\subsection{Quantum fluctuations} 
\label{sec3a}

We have performed spin wave theory of Eq.~\eqref{k1} using the standard approach (see the Appendix).\cite{kle1, pw, jon, joli}  The crucial feature of  the Hamiltonian, Eq.~\eqref{k1}, is that the average spin-deviation operator $\la n_{l\alpha}\ra=\la b_{l\alpha}^\dagger b_{l\alpha}\ra$  is much smaller than the U(1)-invariant model.
\begin{figure}[ht]
\centering
\includegraphics[width=3in]{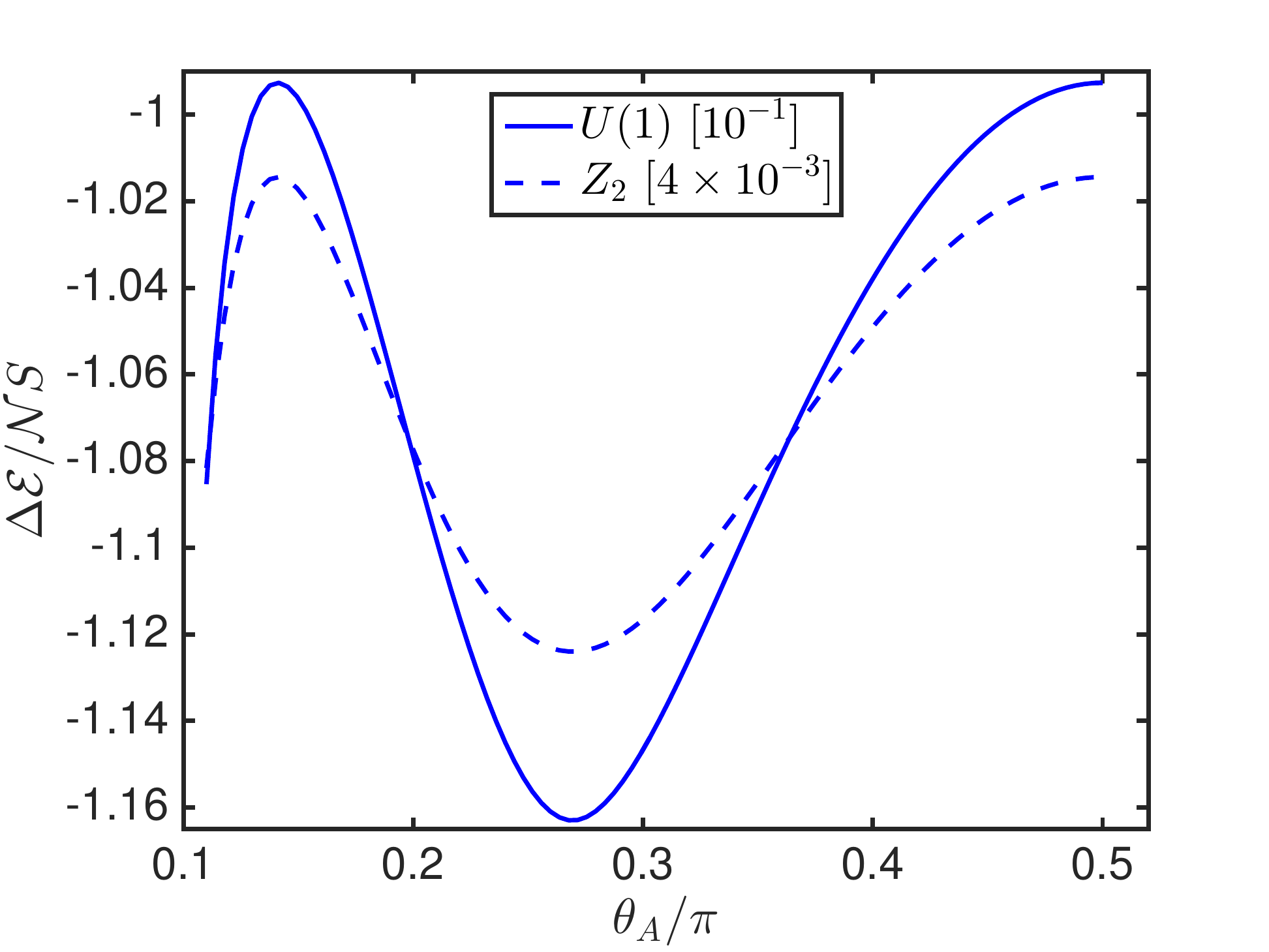}
\caption{The zero point quantum correction to the classical ground state as a function of $\theta_A$ for the U(1)-invariant XXZ model and the $Z_2$-invariant XXZ model at $J_{\pm\pm}/J_z=0.3$; $h=0$. }
\label{Ground_corr}
\end{figure}
\begin{figure}[ht]
\centering
\includegraphics[width=3in]{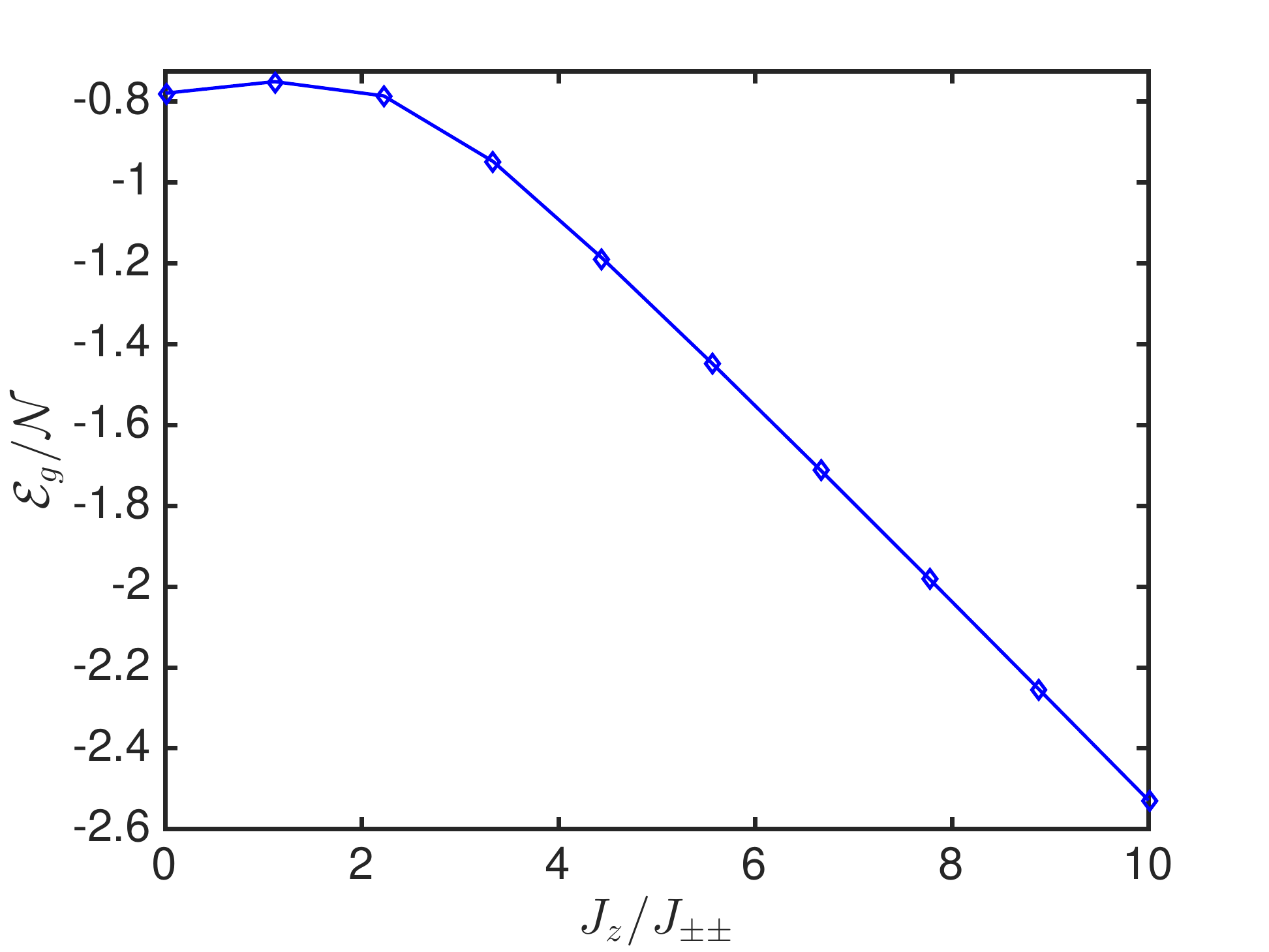}
\caption{The spin wave ground state energy per site  as a function of $J_z/J_{\pm\pm}$ for the $Z_2$-invariant XXZ model at $h=0$, and $S=1/2$. }
\label{GS}
\end{figure}
As we shall see in the subsequent sections, the physical reason behind this is because the excitation spectra of  Eq.~\eqref{k1} are gapped in the entire Brillouin zone, with a maximum contribution from the $\bo_\Gamma=0$ mode. As a result all the points in the Brillouin zone are finite. Since  $\la n_{l\alpha}\ra\propto 1/\epsilon_{\bo\alpha}$, the smallness of this quantity comes from the $\bo_\Gamma=0$ mode.   Thus, quantum fluctuations are suppressed in the $Z_2$-invariant model and spin wave theory is suitable for describing  the ground state properties of the {\it quantum triangular ice} in Eq.~\eqref{k1}.

To corroborate these facts, in Fig.~\eqref{Ground_corr} we show the quantum correction to the classical energy as a function of the one-parameter degeneracy $\theta_A$ for   the U(1)-invariant model\cite{kle1} and the $Z_2$-invariant model in Eq.~\eqref{k1}. In both cases, quantum fluctuations select a particular $\theta_A$ ground state. As the minimum energy occurs at the same value of $\theta_A$ in both models, the selected $\theta_A$ is given by \cite{kle1} $\cos^2\theta_A=(1-2\kappa)/(1-\kappa^2)$ in both cases. It is easy to verify that $\theta_A=\theta_B\neq \theta_C$ at the true ground state.  However, we also see that the roles of quantum fluctuations  are markedly different in both models. The energy correction in the $Z_2$-invariant model is much smaller than the U(1)-invariant model up to a negative sign. This is in accordance with the suppression of quantum fluctuations.  The ground state energy of the  $Z_2$-invariant XXZ model in the large $J_z$ limit is shown in Fig.~\eqref{GS}, with a maximum peak at the rotationally symmetric point $J_{\pm\pm}=J_z=1;~h=0$ (see Sec.~\ref{cfm}). 

\subsection{ Excitation spectra } 
\label{sec5}
In this section, we address the nature of the $Z_2$ quantum fluctuations  by studying the energy dispersion in different phases. We set $J_z$ as the unit of energy.  At $h=0$, the easy-axis ferromagnetic phase is dominant in the regime $J_{\pm\pm}/J_z>0.5$, which becomes canted for $h<h_F$. In this phase there is only one sublattice on the triangular lattice. All the thermodynamic quantities of interest can be computed analytically as presented in Sec.~\eqref{cfm}. 
\begin{figure}[ht]
\centering
\includegraphics[width=3in]{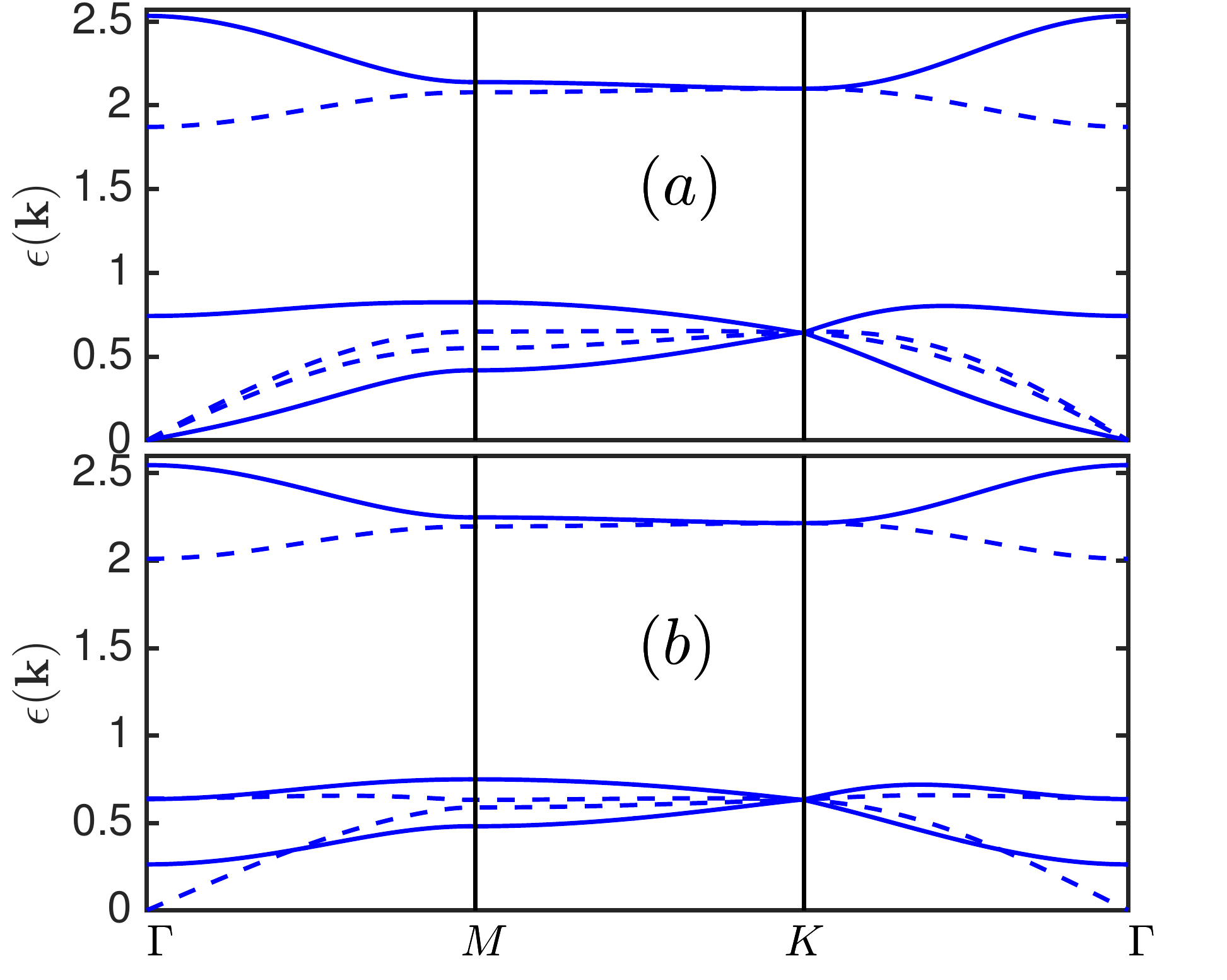}
\caption{The excitation spectra of the U(1) supersolid phase (dashed line)  and the $Z_2$ {\it ferrosolid} phase (solid line) at $J_{\pm\pm}/J_z=0.3$; $h=0$ (a). $J_{\pm\pm}/J_z=0.3$; $h=0.5$ (b).}
\label{FS}
\end{figure}

Now, consider the {\it ferrosolid} phase which is predominant in the regime  $J_{\pm\pm}/J_z<0.5$ with $h=0$ or $h<h_N$. 
In Fig.~\eqref{FS} we have shown the energy spectra in the  supersolid phase of the U(1)-invariant model and the {\it ferrosolid} phase of the $Z_2$-invariant model at $J_{\pm\pm}/J_z=0.3$, $h=0$ (a) and $h=0.5$ (b). The supersolid phase is well-known.\cite{kle1} In this case, there are two gapless modes, one is the degenerate mode, the other is the $S_z$ mode, the gapped mode  is the optical mode. All the modes exhibit a minimum value at $\bo_\Gamma=0$ with linearly dispersing sound modes near this point. 
\begin{figure}[ht]
\centering
\includegraphics[width=3in]{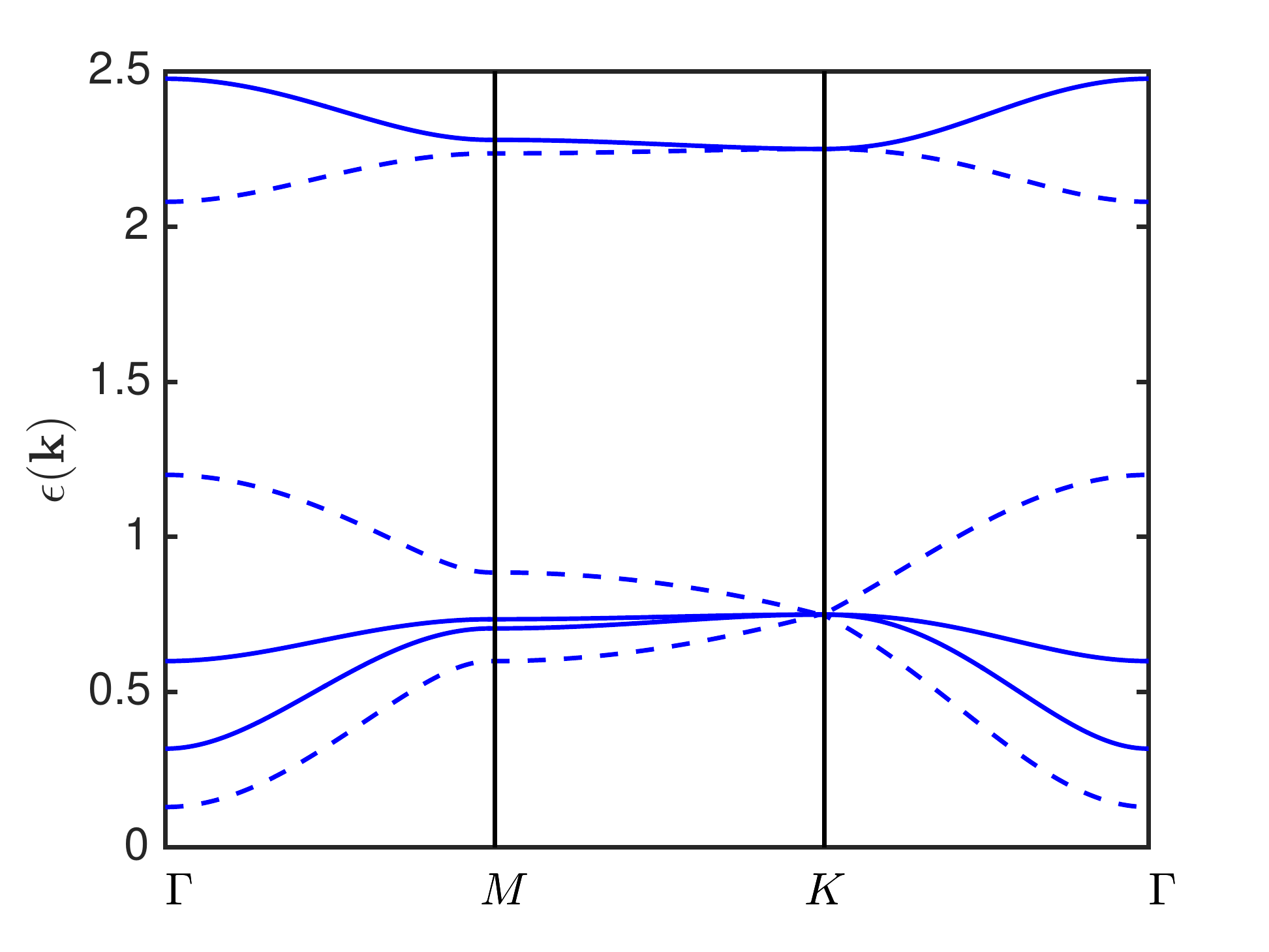}
\caption{The excitation spectra inside the $m=1/6$ lobe  for the U(1)-invariant model (dashed lines) and  the $Z_2$-invariant model (solid lines) at $J_{\pm\pm}/J_z=0.3$; and $h=1.5$.}
\label{NS}
\end{figure}

In contrast, the excitation spectra of the  {\it ferrosolid} phase differs by a gap $S_z$ mode as $S_z$ is not conserved.  The optical mode is gapped, but it comes with a maximum excitation at $\bo_\Gamma=0$. The degenerate mode (in linear spin wave theory) is always gapless  at $\bo_\Gamma=0$ and it comes with linear dispersing sound mode.     As shown in Fig.~\eqref{FS} (b), the  degenerate modes in both  models are lifted by a nonzero magnetic field. However, the $S_z$ mode in the supersolid phase remains gapless at nonzero field since the total $S_z$ is still conversed in this case.  Note that the value of the energy spectra in both models is the same at the corners of the Brillouin zone (that is the $K$ point and the symmetry related points). In other words, the physics of both models are immutable at the corners of the Brillouin zone. Figure~\eqref{NS} shows the excitation spectra inside the classical $m=1/6$ lobe.  In this lobe, the excitation spectra are always gapped in both models. However, we see that the top band  in the $Z_2$-invariant model has its maximum at the $\bo_\Gamma=0$ mode.  Similar to previous phases, the energy bands in both models touch each other at the corners of the Brillouin zone.

\subsection{  Order parameters}
\label{sec7}
To understand the role of quantum fluctuations in the $Z_2$-invariant XXZ model, we now study the effects of the $Z_2$-invariant quantum fluctuations on the order parameters in the large-$S$ expansion.  By means of the generalized Bogoliubov transformation (see the Appendix), the ground state sublattice magnetizations can be computed numerically. They are given by
\begin{align}
\la \bold{S}_\alpha\ra =(SN-\sum_\bo \sum_{n=4}^6 |\mathcal{U}_{m,n}|^2)\hat{\bold{m}}_\alpha,
\label{maga}
\end{align}
where $\mathcal{U}$ is the an $6\times 6$ matrix that diagonalizes the spin wave Hamiltonian (see the Appendix for more details), $m= 1, 2, 3$ for the sublattice $\alpha=A,B,C$ respectively.  
\subsubsection{ Order parameters at zero magnetic field }
\begin{figure}[ht]
\centering
\includegraphics[width=3in]{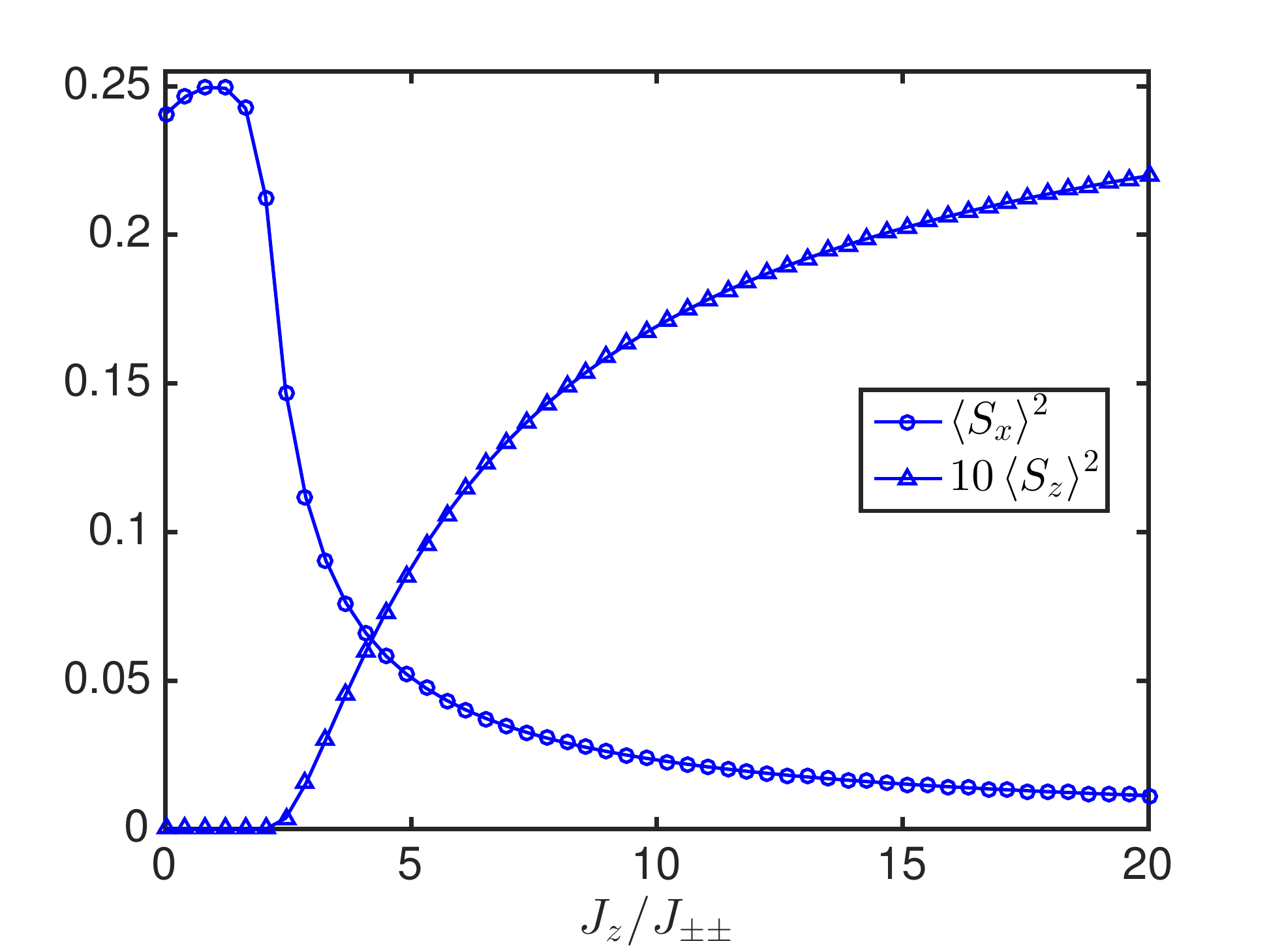}
\caption{The plot of the total $S_x$ and $S_z$ order parameters squared as a function of $J_z/J_{\pm\pm}$ for the $Z_2$-invariant XXZ model at $h=0$ and $S=1/2$.}
\label{ST}
\end{figure}

 At $h=0$, Eq.~\eqref{k1} has a one-parameter family of degenerate ground states, which are lifted by quantum fluctuations.  There are two possible quantum phases in the $h=0$ limit. $(i)$ The {\it ferrosolid} state,  which is selected by quantum fluctuations in the frustrated regime $J_z\gg J_{\pm\pm}$. $(ii)$ The $x$-easy-axis ferromagnetic ordered state resulting from the spontaneous $Z_2$ symmetry breaking along the $x$-axis of the spin. This phase appears in the unfrustrated regime, $J_{\pm\pm}\gg J_z$ (see Sec.~\ref{cfm}).  The system is characterized by two order parameters,  $\la S_x\ra$ and  $\la S_z\ra$.   Figure~\eqref{ST} shows the order parameters as a function of $J_z/J_{\pm\pm}$ at $h=0$. We see that the $\la S_x\ra$ order parameter displays an exact maximum peak at the rotationally symmetric point $J_z/J_{\pm\pm}=1$,  with $\la S_x\ra_{\text max}=S=0.5$ (see Sec.~\ref{cfm}), and decreases monotonically with increasing $J_z/J_{\pm\pm}$, but never vanishes for any finite values of $J_z/J_{\pm\pm}$.  The survival of the $S_x$-order parameter at arbitrary large $J_z/J_{\pm\pm}$ signifies that the system still exhibits spontaneously broken $Z_2$ symmetry. This is accompanied by a simultaneous non-vanishing Ising order at arbitrary large $J_z/J_{\pm\pm}$. The $\la S_z\ra$ order parameter is completely zero in the $x$-easy-axis ferromagnetic phase, which appears for $J_z/J_{\pm\pm}\leq 2$ (see Sec.~\ref{cfm}). In the dominant $J_z/J_{\pm\pm}>2$ limit,  $\la S_z\ra$ increases with increasing $J_z/J_{\pm\pm}$.  The system is then characterized by two non-vanishing order parameters,  $\la S_z\ra\neq 0$ and  $\la S_x\ra\neq 0$, which is  a {\it ferrosolid} state. 

As in bosonic systems, there is a continuous quantum phase transition connecting the regime of small $J_z$ (easy-axis ferromagnet) and that of large $J_z$ ({\it ferrosolid}). The dynamical static structure factors $\mathcal{S}^{\pm}(\bo)$ and  $\mathcal{S}^{zz}(\bo)$ can be related to the squares of the order parameters in the thermodynamic limit. We find that in addition to the usual Bragg peaks at the corners of the Brillouin zone for moderate values of  $J_z/J_{\pm\pm}$, there is a  non-divergent contribution at $\bo_\Gamma$.  At the rotationally symmetric point, $\mathcal{S}^{\pm}(\bo)$ is zero and  $\mathcal{S}^{zz}(\bo)$ is nonzero but completely flat (see Sec.~\ref{cfm}). These two quantities are divergent at $\bo_\Gamma$ in the U(1)-invariant XXZ model due to zero energy contributions.

\subsubsection{ Order parameters at nonzero magnetic field }
\begin{figure}[ht]
\centering
\includegraphics[width=3in]{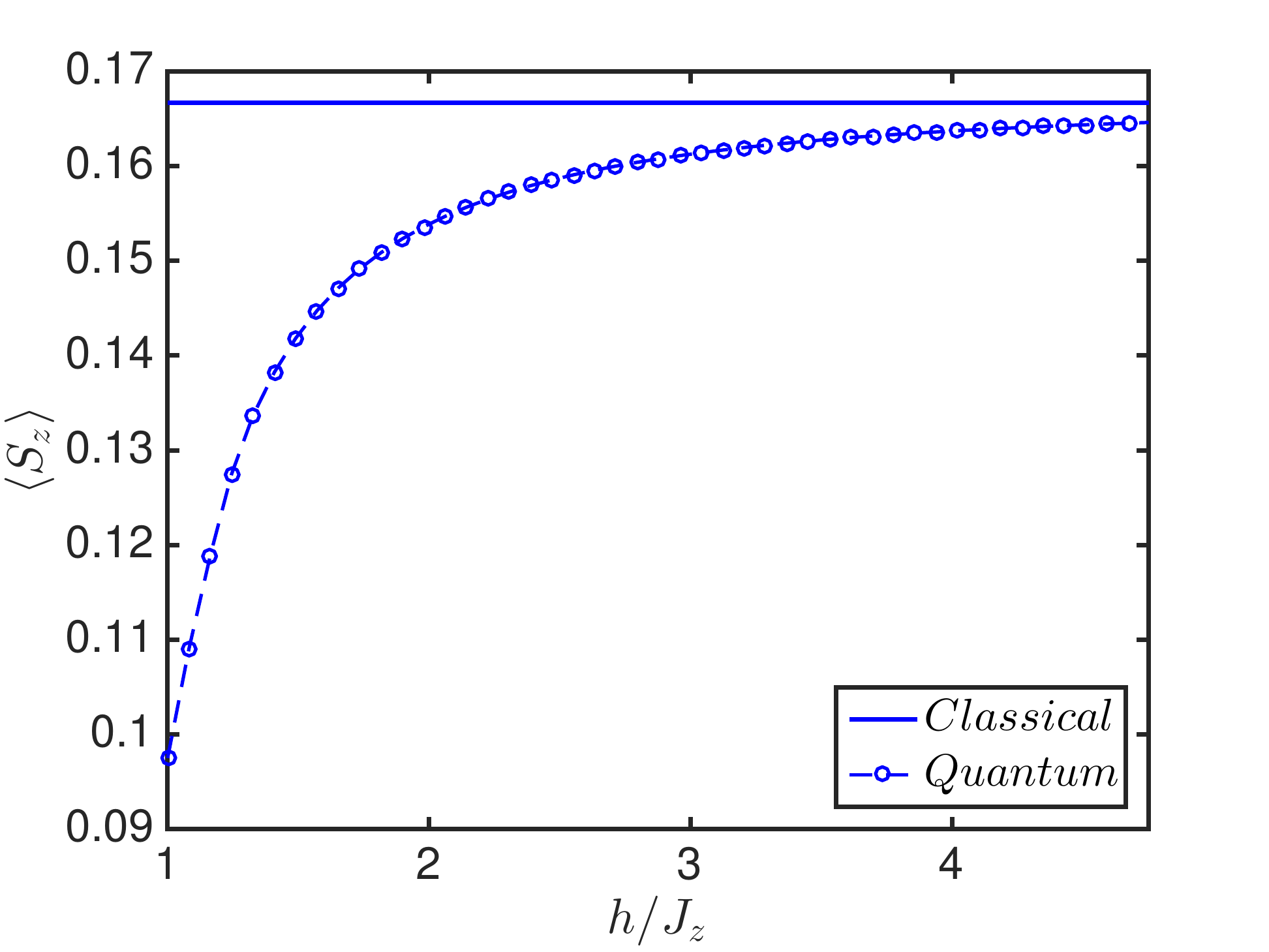}
\caption{The plot of the classical and quantum  total $S_z$ against the magnetic field inside the $m=1/6$ lobe for the $Z_2$-invariant XXZ model at $S=1/2$,  $J_{\pm\pm}/J_z=0.3$. }
\label{Z2_lobe1}
\end{figure}
\begin{figure}[ht]
\centering
\includegraphics[width=3in]{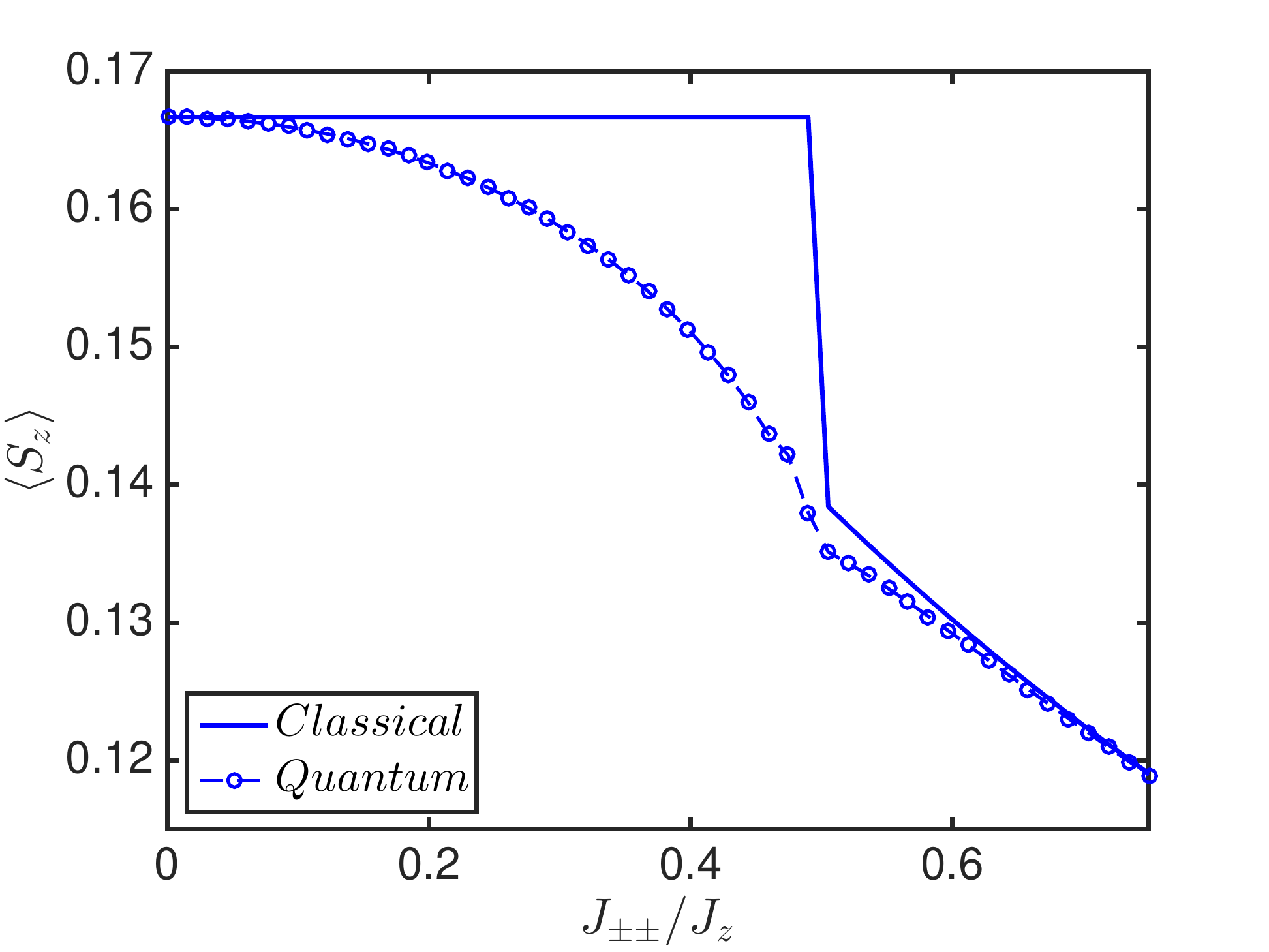}
\caption{The plot of the classical and quantum  total $S_z$ against the anisotropy entering the $m=1/6$ lobe for the $Z_2$-invariant XXZ model at $S=1/2$, and $h/J_z=2.5$. }
\label{Z2_lobe2}
\end{figure}
Now, we study the nature of the classical $m=\pm 1/6$ lobes which appear for nonzero magnetic field and $J_{\pm\pm}/J_z<0.5$.   In the U(1)-invariant model, \cite{fa, fa1,has0,has,has1, has2,kle1,isa, isa1,isa2,isa3, roger, xu,dar} the classical $m=\pm 1/6$ are unchanged by quantum fluctuations. In other words, they are fully polarized Mott phases.  They are incompressible and exhibit zero magnetic susceptibility. In the present model,  the classical $m=\pm 1/6$ lobes  are altered by the $Z_2$ quantum fluctuations and they  retain a small finite magnetic susceptibility as depicted in Figs.~\eqref{Z2_lobe1} and \eqref{Z2_lobe2}. Hence, the $S_z$ or the Ising order characterizing magnetic ordering inside the lobes is destroyed, and the $\la S_x\ra$-order parameter remains zero inside the lobes. This is  similar to the ice-like property observed from the QMC simulations on the kagome lattice,\cite{juan,sow1} which appears in the regime $J_{\pm\pm}/J_z<0.5$; $h/J_z\neq 0$, and promote a putative  QSL state on the kagome lattice. The absence of order leads to diffuse peaks in the dynamical structure factors at $\bo_\Gamma$, as well as the corners of the Brillouin zone.



\section{Easy-axis CFM}
In this section, we explore the ground state properties of the easy-axis canted ferromagnetic phase of Eq.~\eqref{k1}. A related U(1)-invariant XXZ model in the superfluid phase has been studied previously on the triangular lattice, using series expansion methods\cite{AW} and spin wave theory. \cite{AW1} The present model has not been studied in the ferromagnetic phase. The understanding of this phase might be applicable to many systems,   since many magnetic materials are ferromagnets.  The ferromagnetic coupling also plays a prominent role in the experimental realization  of hard-core bosons in ultracold atoms on quantum optical lattices.\cite{bec,mar}  We consider a more general Hamiltonian given by
\begin{align}
&H=J_{\pm\pm}\sum_{\la lm\ra}\lb  -S_{l}^xS_{m}^x +S_{l}^yS_{m}^y\rb + \sum_{lm}J_{l,m}  S_{l}^zS_{m}^z \nonumber\\&-h_z\sum_l S_l^z-h_x\sum_l S_l^x,
\label{kk}
\end{align}
where $J_{l,m}=J_z$ on the nearest-neighbour (nn) sites and $J_{l,m}=J^{\prime}_z$ on the next-nearest-neighbour (nnn) sites.  The external magnetic fields are introduced to enable the calculation of parallel and longitudinal magnetizations. \label{cfm} This model retains $Z_2$ symmetry when $h_x=0$.

\subsection{Spin wave theory for the easy-axis CFM }
\label{class}
We first   consider the rotationally symmetric point $J_z^\prime=h_{x,z}=0$ and $J_z=J_{\pm\pm}=1$.   The resulting Hamiltonian  in this limit can be written as
\begin{align}
&H=\sum_{\la lm \ra} -S_{l}^xS_{m}^x +S_{l}^yS_{m}^y+S_{l}^zS_{m}^z,\nonumber\\&
=\sum_{\la lm \ra} -S_{l}^xS_{m}^x +\frac{1}{2}(S_{l}^+S_{m}^- + S_{l}^-S_{m}^+).
\label{spl}
\end{align}
In the last line, we have chosen $x$-quantization axis, hence $S^\pm = S^z\pm iS^y$.   The system exhibits spontaneously broken $Z_2$ symmetry along the $x$-axis as  opposed to the spontaneously broken U(1) symmetry along the $x$-$y$ plane in the U(1)-invariant XXZ model.\cite{fa, fa1,has0,has,has1, has2,kle1,isa, isa1,isa2,isa3, roger, xu,dar} The state $\ket{\psi_{FM}}=\prod_l\ket{S_{l}^x=\uparrow}$  is an exact eigenstate of Eq.~\eqref{spl} and the corresponding energy is  $\mathcal{E}_{g}/\mathcal{N}=-3S^2=-0.75$ for $S=1/2$, which corresponds to the maximum energy peak in Fig.~\eqref{GS}. The order parameter is, $\la S_x\ra=S$. Thus, spin wave theory is exact in this limit and the ground state is the $x$-easy-axis ferromagnetic phase whose spectrum is gapped at all excitations, as will be shown in the subsequent sections.

For the anisotropic model, the mean field energy in the ferromagnetic phase is given by
\begin{align}
&\mathcal{E}_{MF}(\theta)= 3\mathcal{N}S^2( \lambda_{1z}+ \lambda_{2z}-{h}/{3}),
\label{emf}
\end{align}
where $\mathcal{N}$ is the total number of sites and
\begin{align}
&\lambda_{1z}= J_z\cos^2\theta-J_{\pm\pm}\sin^2\theta;\quad\lambda_{2z}= J_z^\prime\cos^2\theta;
\\&h=h_z\cos\theta +h_x\sin\theta. 
\end{align}
The minimization of the mean field energy with respect to $\theta$ yields
\begin{align}
h_z-6\mathcal J \cos\vartheta=h_x\cot\vartheta,
\label{hh1}
\end{align}
where $\vartheta$ is the angle that minimizes Eq.~\eqref{emf} and $\mathcal J=\lb J_z+J_z^\prime+J_{\pm\pm} \rb$. We can eliminate $h_x$ from Eq.~\eqref{emf} using Eq.~\eqref{hh1},  the classical energy becomes
\begin{align}
\mathcal{E}_{MF}(\vartheta)= 3\mathcal{N}S^2[\mathcal J \sin^2\vartheta + J_z+J_z^\prime-h_z\sec\vartheta/3].
\end{align}
It should be noted that $\vartheta$ is a function of $h_x$. 
At $h_x=0$, the solution of Eq.~\eqref{hh1} gives $\cos\vartheta_0=h_z/h_c$, where $h_c=6\mathcal J$ is the critical field above which the spins  are fully polarized along the $z$-direction. The corresponding classical energy is given by
\begin{align}
\mathcal{E}_{MF}(\vartheta_0)= -3\mathcal{N}S^2[\mathcal J \cos^2\vartheta_0+J_{\pm\pm}].\label{mf2} 
\end{align}
For $h_x\neq 0$, the mean field energy is obtained perturbatively for small $h_x$,
\begin{align}
\mathcal{E}_{MF}(\vartheta)=\mathcal{E}_{MF}(h_x=0)+ h_x\left.\frac{\partial \mathcal{E}_{MF}(\vartheta)}{\partial h_x}\right\vert_{h_x=0} +\cdots,
\end{align}
where $\mathcal{E}_{MF}(h_x=0)=\mathcal{E}_{MF}(\vartheta=\vartheta_0)$.

We perform  spin wave theory in the usual way, by rotating the coordinate about the $y$-axis in order to align the spins along the selected direction of the magnetization. 
\begin{align}
&S_l^x=S_l^{\prime x}\cos\theta  +  S_l^{\prime z}\sin\theta,\label{trans}\nonumber\\&
S_l^y=S_l^{\prime y},\\&\nonumber
S_l^z=- S_l^{\prime x}\sin\theta + S_l^{\prime z}\cos\theta.
\end{align}
As mentioned above, there is only one sublattice in this phase.  Next, we express the rotated coordinates in terms of the linearized Holstein Primakoff (HP) transform.

 \begin{align}
 &S_{l}^{\prime z}= S-b_{l}^\dagger b_{ l}, \label{HP}\nonumber\\&
 S_{l}^{\prime y}= i\sqrt{\frac{S}{2}}\lb b_{l}^\dagger -b_{l}\rb,
 \\&\nonumber
 S_{l}^{\prime x}= \sqrt{\frac{S}{2}}\lb b_{l}^\dagger +b_{l}\rb.
 \end{align}
 
 The truncation of the HP transformation at linear order is guaranteed provided the average spin-deviation operator $\la n_l\ra=\la b_l^\dagger b_l\ra$ is small. Indeed,  $\la n_l\ra$ is small in the present model, as presented above.
Taking the magnetic fields, $h_{x,z}$, to be of order $S$ and keeping only the quadratic terms,
 the resulting bosonic Hamiltonian can be diagonalized by the Bogoliubov transformation,
\begin{align}
b_{\bo}=u_{\bo}\alpha_{\bo}-v_{\bo}\alpha_{-\bo}^\dagger,
\end{align}
where $u_{\bo}^2-v_{\bo}^2=1$, one finds that the resulting Hamiltonian is diagonalized by
\begin{align}
&u_{\bo}^2=\frac{1}{2}\lb \frac{A_{\bo}(\vartheta)}{\omega_{\bo}(\vartheta)}+1\rb; \quad v_{\bo}^2=\frac{1}{2}\lb \frac{A_{\bo}(\vartheta)}{\omega_{\bo}(\vartheta)}-1\rb,\end{align}
with $\omega_\bo(\vartheta)=\sqrt{A_\bo^2(\vartheta)-B_\bo^2(\vartheta)}$.  The diagonal Hamiltonian yields
\bea
H=S\sum_{\bo}\omega_{\bo} (\vartheta)\lb \alpha_{\bo}^\dagger \alpha_{\bo}+\alpha_{-\bo}^\dagger \alpha_{-\bo}\rb.
\eea

The excitation of the quasiparticles is given by

\bea
\epsilon_\bo(\vartheta)=2S\omega_\bo(\vartheta)=2S\sqrt{A_\bo^2(\vartheta)-B_\bo^2(\vartheta)},
\eea
while the spin wave ground state energy is given by \bea \mathcal{E}_{SW}(\vartheta)=\mathcal{E}_{MF}(\vartheta) +S\sum_{\bo}[\omega_{\bo}(\vartheta)-\mathcal{E}_{lo}(\vartheta)] \label{Esw},\eea
where
\begin{align}
&A_\bo(\vartheta)=\mathcal{E}_{lo}(\vartheta) +3J_{\pm\pm}\gamma_{\bo}+ B_\bo(\vartheta);\\& B_\bo(\vartheta)=\frac{3}{2}\lb \bar{g}_\bo(\vartheta)+g_\bo(\vartheta)\rb;
\end{align}

\begin{align}
&\mathcal{E}_{lo}(\vartheta)=-3(J_{z}+J_{z}^\prime)+h_z\sec\vartheta/2,\\&
\bar{g}_\bo(\vartheta)=(J_z+J_{\pm\pm})\sin^2\vartheta\gamma_{\bo}-2J_{\pm\pm}\gamma_{\bo},\\&g_\bo(\vartheta)=J_z^\prime\sin^2\vartheta\bar{\gamma}_\bo.
\end{align}
We have eliminated $h_x$ using Eq.~\eqref{hh1}.
The structure factors are given by
\begin{align}
&\gamma_{\bo}=\frac{1}{3}\lb \cos k_x+2\cos \frac{k_x}{2}\cos \frac{\sqrt{3}k_y}{2}\rb,\\&
\bar{\gamma}_{\bo}=\frac{1}{3}\lb \cos \sqrt{3}k_y+2\cos \frac{3k_x}{2}\cos \frac{\sqrt{3}k_y}{2}\rb.
\end{align}

\subsection{Excitation spectra for the easy-axis CFM}
\begin{figure}[ht]
\centering
\includegraphics[width=3in]{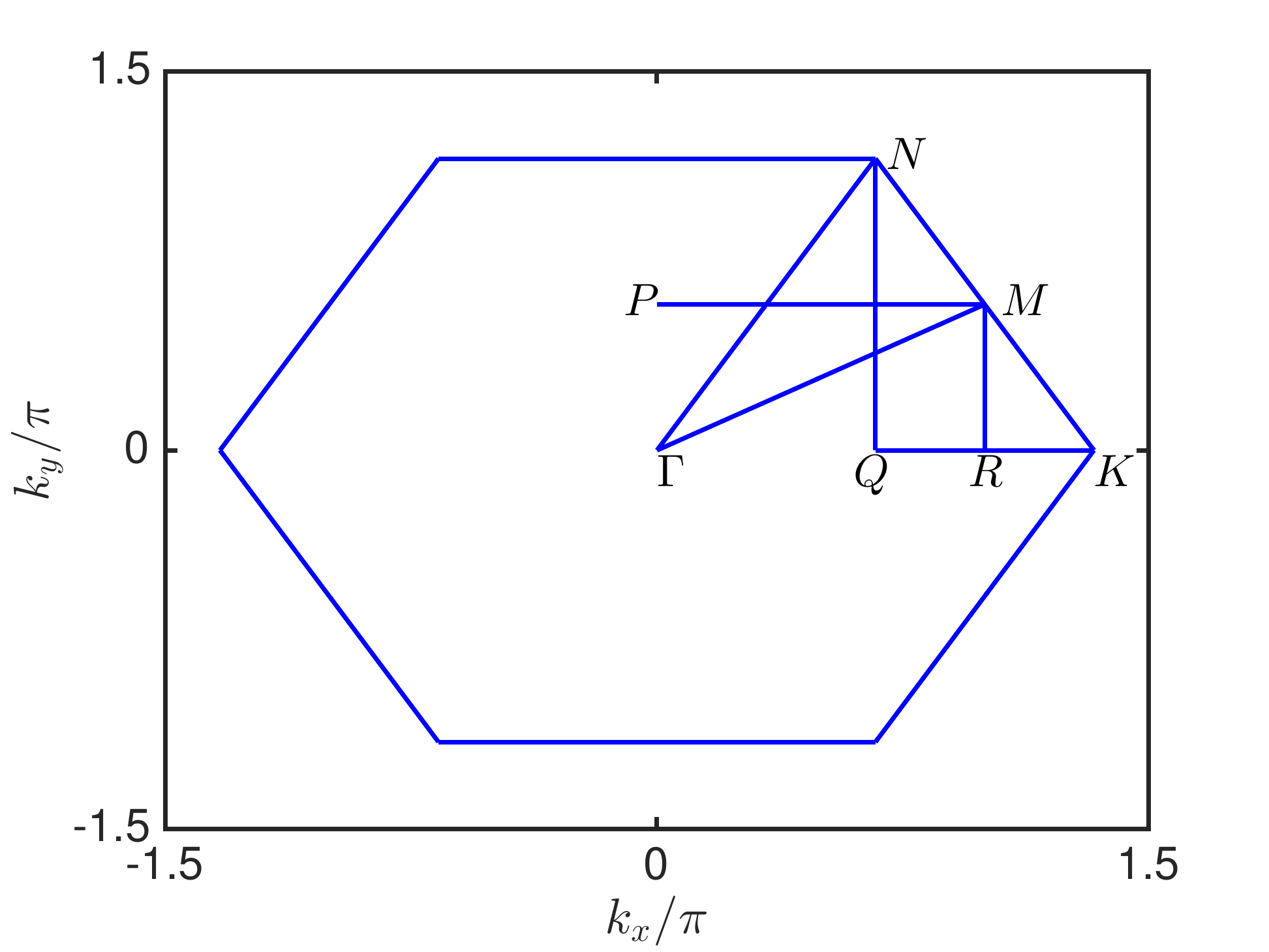}
\caption{Color online.  The full Brillouin zone of the triangular lattice and the corresponding paths that will be adopted in this section.}
\label{K_spec1}
\end{figure}

\begin{figure}[ht]
\centering
\includegraphics[width=3in]{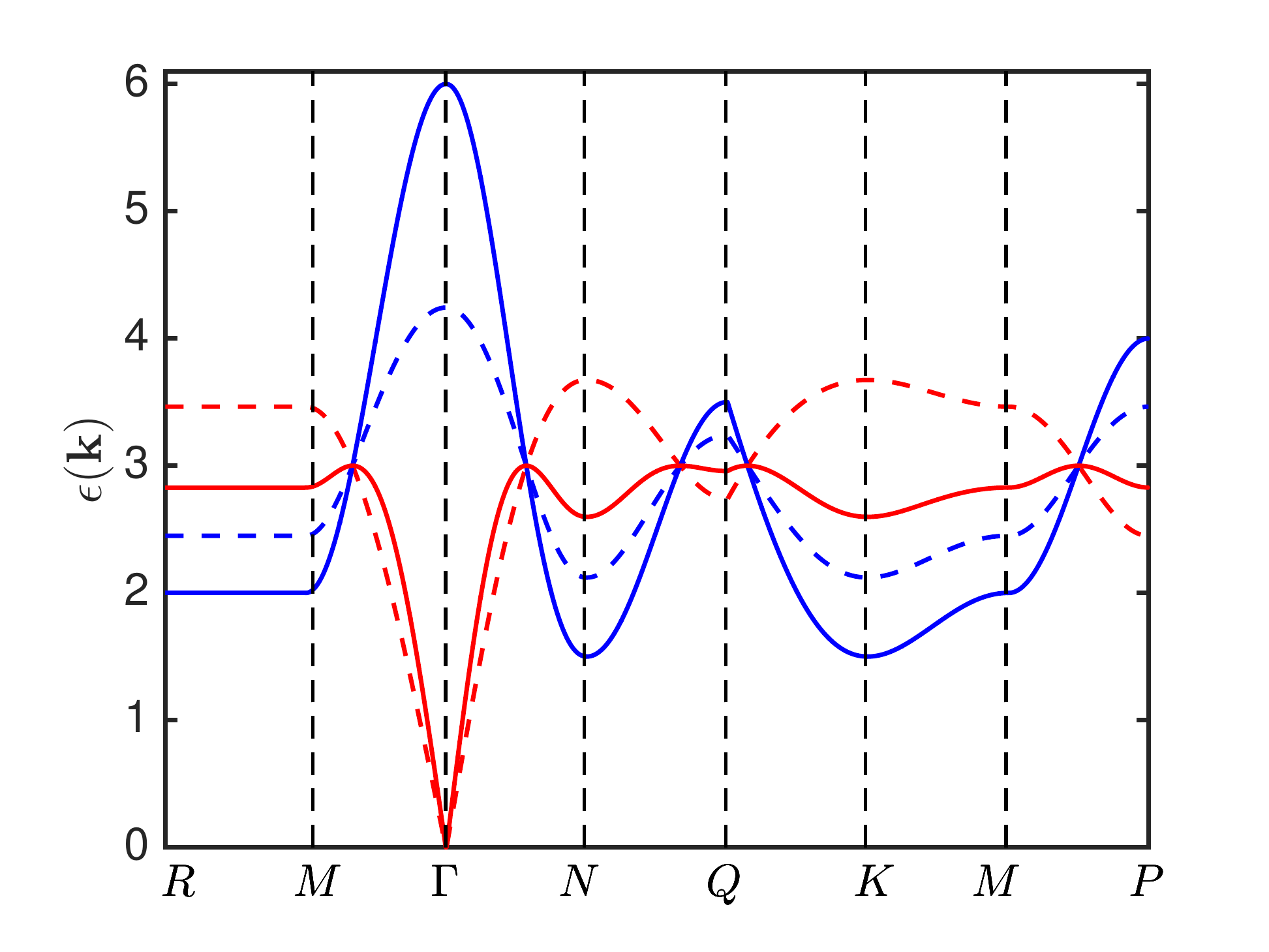}
\caption{Color online. The plots of the energy dispersion  at $h_{x,z}=0~(\rho=0.5)$ and $J_{\pm\pm}=1;~ J_z^\prime=0$. XY model: $J_z=0$  (dashed). Heisenberg model:  $J_z =1$ (solid). The blue curves denote the present model and the red curves denote the U(1) model.}
\label{ee_nn0}
\end{figure}

 We now investigate the nature of the energy spectra of Eq.~\eqref{kk}. We will be interested in half-filling or zero magnetic fields  and spin-$1/2$. We adopt the Brillouin zone paths of Refs.~[\onlinecite{AW, AW1}] as shown in Fig.~\eqref{K_spec1}. Our main focus is the appearance of a minimum inside the Brillouin zone (roton minimum) and the possibility of any soft modes. The vanishing of the spectrum at the corners of the Brillouin zone represents a phase transition to a new spin configuration.  The simplest case is the rotationally symmetric point  $h_{x,z}=J_z^\prime=0;~ J_z=J_{\pm\pm}=J$. As shown above, spin wave theory is exact in this limit  and the excitation spectrum of Eq.~\eqref{spl} is $\omega_\bo=A_\bo= 3J(1+\gamma_\bo)$, where  $B_\bo=0$. We see that the exact ground state is the fully polarized easy-axis ferromagnet along the $x$-axis, and the corresponding excitation exhibits no zero modes since $-1/2\leq \gamma_\bo\leq 1$; see Fig.~\eqref{ee_nn0}. Near $\bo_\Gamma=0$, the energy behaves as $\omega_\bo\approx a-b\bo^2$, with $a=6J$ and $b=3J/4$. As  Mermin-Wagner theorem\cite{mer} does not apply to $Z_2$ invariant Hamiltonian,  the average spin deviation operator near this mode at low-temperature is not divergent,
\begin{align}
\Delta S_x(T)=\int \frac{ d\bo}{e^{\omega_\bo/k_BT}-1}\sim~ T\frac{\ln(1-e^{-a})}{b}.
\label{mw}
\end{align}

\begin{figure}[ht]
\centering
\includegraphics[width=3in]{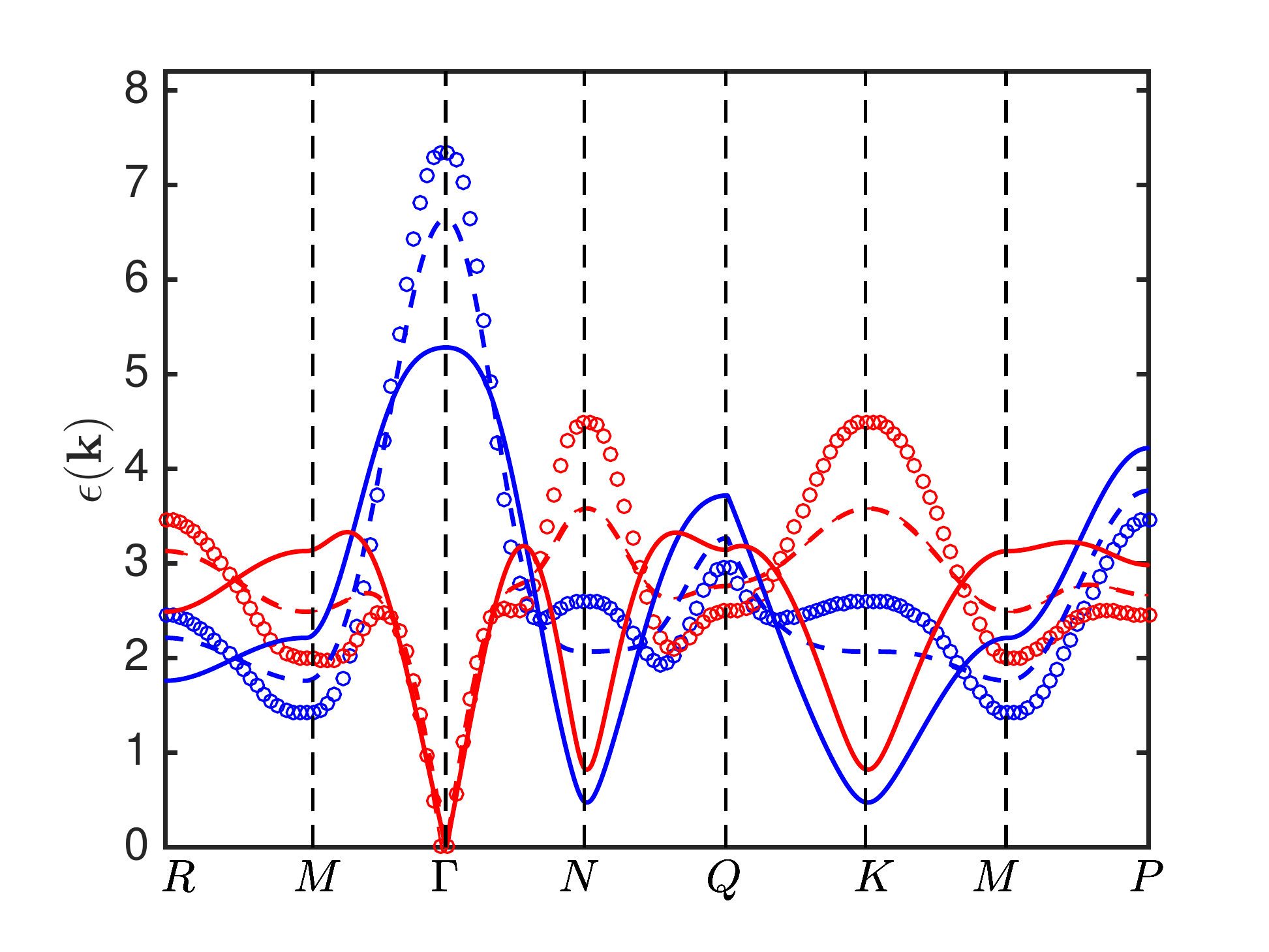}
\caption{Color online. The plots of the energy dispersion  at $h_{x,z}=0~(\rho=0.5)$, $J_{\pm\pm}=J_z=1$ and several values of $J_z^\prime=0.45$ (dashed),  $J_z^\prime =-0.45$ (solid), and  $J_z ^\prime =1$ (symbol). The colors have the same meaning as in Fig.~\eqref{ee_nn0}.}
\label{ee1}
\end{figure}
Figures \eqref{ee_nn0} and \eqref{ee1} show the energy spectra of the $Z_2$-invariant XXZ model in the CFM. As previously mentioned, the spectra have a similar behaviour to the U(1)-invariant XXZ model\cite{AW, AW1} at the corners of the Brillouin zone and along $RM$ except for the case of XY model. At the corners of the Brillouin zone,  $\bold{Q}=(\pm 4\pi/3,0)$,  the spectrum vanishes when $J_z^\prime=J_{\pm\pm}-J_z/2$. Hence at $J_z^\prime=0$, the transition from the easy-axis ferromagnet to a {\it ferrosolid} occurs at $J_z=2J_{\pm\pm}$, which happens to be the same as in the U(1)-invariant model.\cite{kle1}  As we have shown above, spin wave theory gives a better description of the present model as quantum fluctuations are very small.   Besides,  QMC simulations \cite{fa, fa1,has0,has,has1, has2,isa, isa1,isa2,isa3, roger,xu,dar} have shown that the superfluid-supersolid transition occur at $J_z/J_{\pm\pm}\approx 4.5$, which is higher than the one predicted by spin wave theory. In other words, quantum fluctuations play a crucial role in the U(1)-invariant XXZ model. Since quantum fluctuations are suppressed in the $Z_2$-invariant model, it would be interesting to verify if  the transition point $J_z/J_{\pm\pm}=2$ is numerically consistent.   

Similar to previous phases, the spectra of the $Z_2$-invariant model in the CFM is  invariably gapped in the entire Brillouin zone and the spectra display a maximum peak at the $\Gamma$ point ($\bo_\Gamma=0$) as opposed to the usual phonon dispersion in rotationally invariant systems.  As mentioned above, the gapped nature of the $Z_2$-invariant model is as a consequence of the $Z_2$-symmetry of the Hamiltonian. The gap at the $\Gamma$ point behaves as $\Delta\propto \sqrt{J_{\pm\pm}(J_z+J_z^\prime +J_{\pm\pm})}$, which vanishes only for $J_{\pm\pm}=0$.    These trends are slightly  modified away from half-filling.   In the subsequent  section, we will show that the $\bo_\Gamma=0$ mode enhances the estimated values of the thermodynamic quantities. 

\subsection{Particle and condensate densities}
 The effects of the gapped excitations in the preceding section are manifested explicitly in the  ground state thermodynamic quantities. In the CFM, all the thermodynamic quantities can be calculated analytically.  We compute  the magnetizations per site given by\cite{ ber, tom}
\begin{align}
&\la S_z\ra= -\frac{1}{S \mathcal{N}}\frac{\partial \mathcal{E}_{SW}(\vartheta_0)}{\partial  h_z},\label{sz}\\&
\la S_x\ra= -\frac{1}{S \mathcal{N}}\left.\frac{\partial \mathcal{E}_{SW}(\vartheta)}{\partial  h_x}\right\vert_{\vartheta=\vartheta_0}\label{sx}.
\end{align}
Using Eq.~\eqref{Esw} we obtain
 \begin{align}
&\la S_z\ra= S\cos\vartheta_0 + \frac{\cos\vartheta_0}{2\Theta_{0}}\frac{1}{\mathcal{N}}\sum_{\bo}\Theta_{\bo}\sqrt{\frac{{A}_ {\bo}(\vartheta_0)-{B}_ {\bo}(\vartheta_0)}{A_ {\bo}(\vartheta_0)+{B}_ {\bo}(\vartheta_0)}},
\label{sz1}
\end{align}
with $\Theta_\bo = (J_z+J_{\pm\pm})\gamma_\bo + J_z^\prime \bar{\gamma}_\bo$. 

The total density of particles is given by $\rho= S+\la S_z\ra$. To linear order in $h_x$ we find
\begin{align}
&\la S_x\ra= S\sin\vartheta_0 - \frac{\cos^2\vartheta_0}{2\Theta_0\sin\vartheta_0}\frac{1}{{\mathcal{N}}}\sum_{\bo}\Theta_{\bo}\sqrt{\frac{{A}_ {\bo}(\vartheta_0)-{B}_ {\bo}(\vartheta_0)}{{A}_ {\bo}(\vartheta_0)+{B}_ {\bo}(\vartheta_0)}}\label{sx1}\\&\nonumber
-\frac{1}{2\sin\vartheta_0}\frac{1}{{\mathcal{N}}}\sum_{\bo}\bigg[\frac{{A}_ {\bo}(\vartheta_0)}{\omega_ {\bo}(\vartheta_0)}-1\bigg].
\end{align}
\begin{figure}[ht]
\centering
\includegraphics[width=3in]{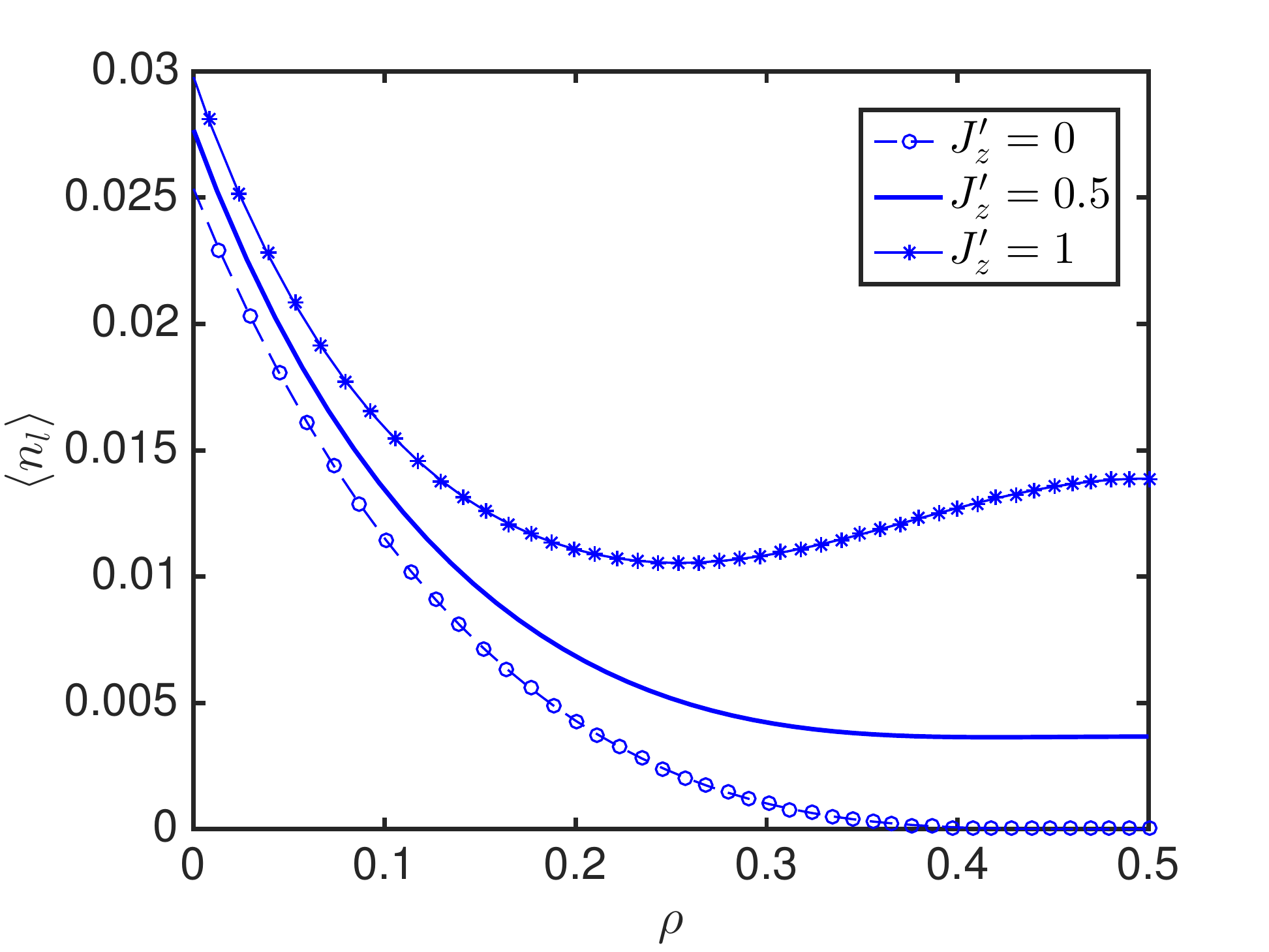}
\caption{Color online. The spin-deviation operator against the particle density at $h_{x}=0$, $J_{\pm\pm}=J_z=1$. }
\label{ave}
\end{figure}
\begin{figure}[ht]
\centering
\includegraphics[width=3in]{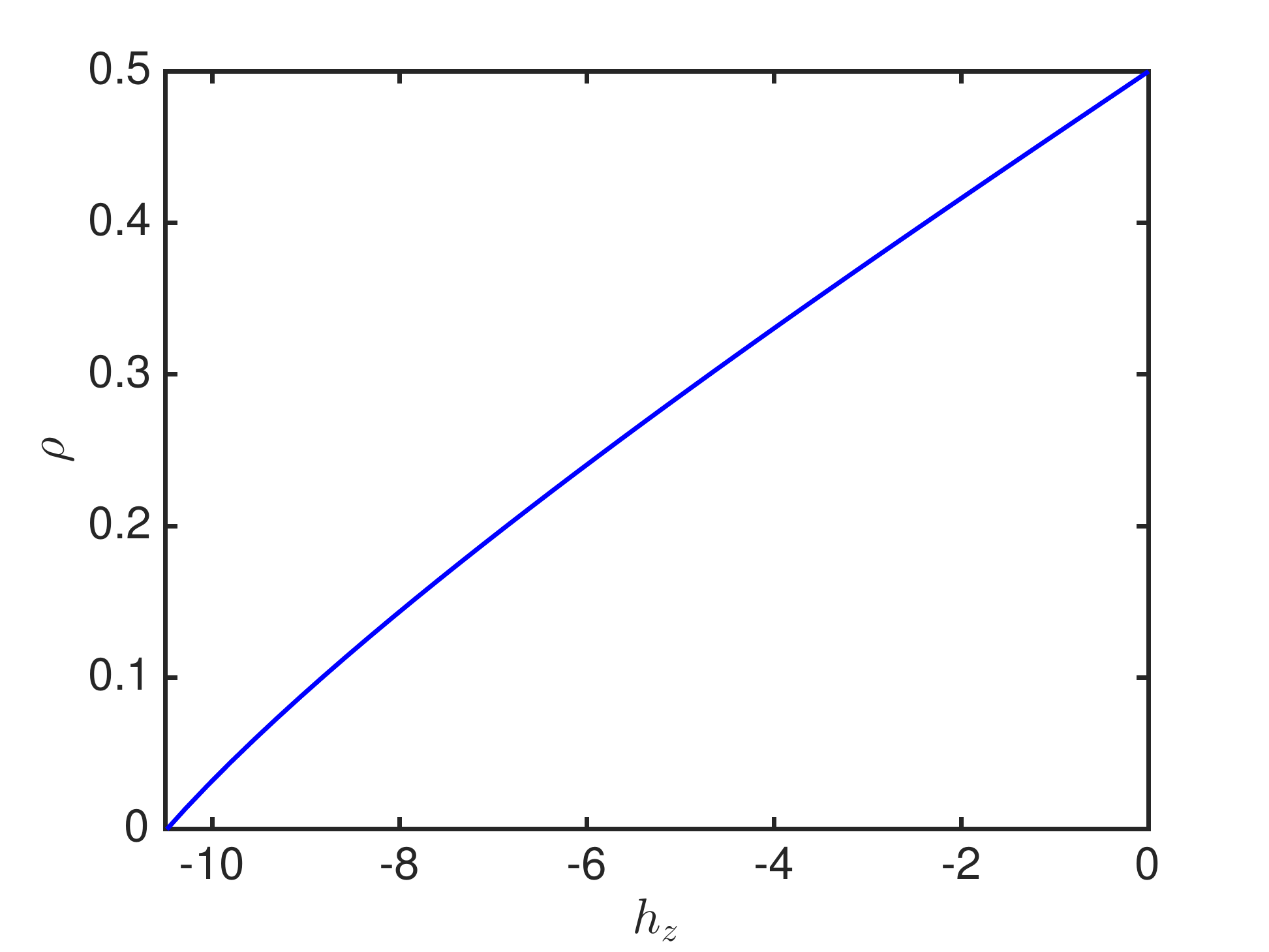}
\caption{Color online. The particle density $\rho$ vs. $h_z$ at $J_z=J_{\pm\pm}=1$, $J_z^\prime=0$, and $S=1/2$.}
\label{rho_hz}
\end{figure} 
Similar to the U(1)-invariant model, the condensate fraction at $\bo_\Gamma=0$ is related to the $S_x$-order parameter by $\rho_0=\lim_{S\to1/2}\la S_x\ra^2$. To linear order in spin wave theory $\rho_0$ is given by
\begin{align}
&\rho_0= \rho_0^c- \frac{\cos^2\vartheta_0}{2\Theta_0}\frac{1}{\mathcal{N}}\sum_{\bo}\Theta_{\bo}\sqrt{\frac{{A}_ {\bo}(\vartheta_0)-{B}_ {\bo}(\vartheta_0)}{{A}_ {\bo}(\vartheta_0)+{B}_ {\bo}(\vartheta_0)}}\label{sx1}\\&\nonumber
-\frac{1}{2}\frac{1}{\mathcal{N}}\sum_{\bo}\bigg[\frac{{A}_ {\bo}(\vartheta_0)}{\omega_ {\bo}(\vartheta_0)}-1\bigg].
\end{align}
where $\rho_0^c =\frac{1}{4}\sin^2\vartheta_0$ is the classical condensate fraction.
\begin{figure}[ht]
\centering
\includegraphics[width=3in]{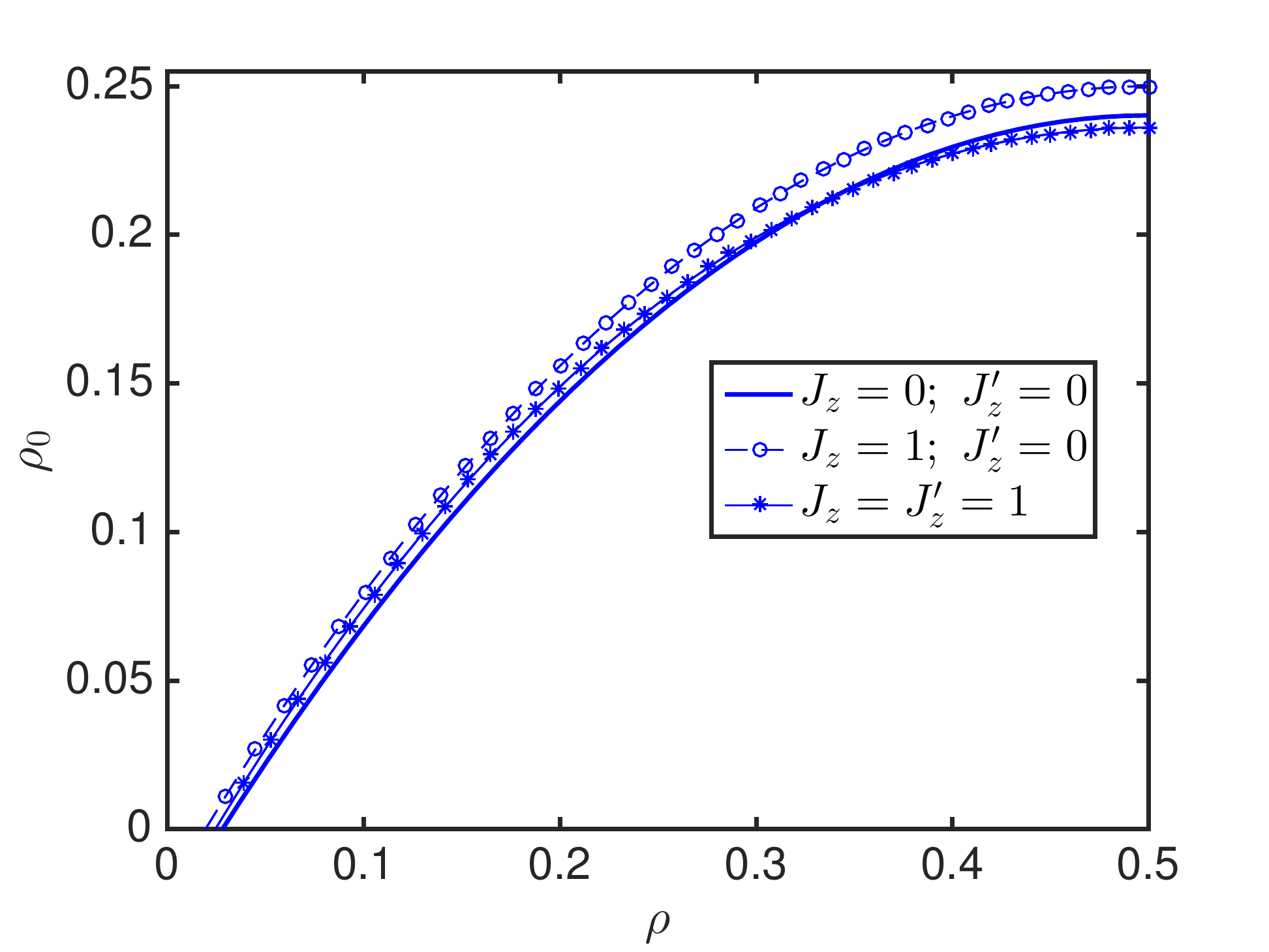}
\caption{Color online. The condensate fraction, $\rho_0$, against the particle density $\rho$ at  $J_{\pm\pm}=1$ and $S=1/2$.}
\label{rho0}
\end{figure}

 An important feature of the $Z_2$-invariant model is that all the thermodynamic quantities are finite at all points in the Brillouin zone. This enhances the estimated values of the thermodynamic quantities.  At the XY point,  $J_z=J_z^\prime=h_{x,z}=0$, the estimated value of the order parameter is $\la S_x\ra= S-0.00981$,\cite{sow3} which should be compared to the O(2)-invariant model $\la  S_x\ra= S-0.05146$.\cite{zhe} A detail analysis of the XY model has been given elsewhere for the triangular and the kagome lattices.\cite{sow3} To further substantiate previous claim of quantum suppression, Fig.~\eqref{ave} shows  the spin-deviation operator $\la n_l\ra$. We see that   $\la n_l\ra$ is extremely small close to half-filling and also very small away from half-filling. In other words, linear spin theory is very suitable for describing the ground state properties of the $Z_2$-invariant XXZ model on non-bipartite lattices.  We have shown the trend of the particle density in Fig.~\eqref{rho_hz} as a function of $h_z$.    In Fig.~\eqref{rho0}, we plot the condensate fraction at $\bo_\Gamma=0$ as a function of $\rho$.  At $\rho=0.5$, the estimated values of $\rho_0$ at the XY point are $\rho_0=0.2402$ linear spin wave theory of the $Z_2$-invariant model,\cite{sow3} $\rho_0=0.2011$ series expansion of the O(2)-invariant model,\cite{zhe} and $\rho_0=0.19127$ second-order spin wave theory on the square lattice. \cite{tom} At the  rotationally symmetric point,  $\rho_0=\rho_0^c=0.25$ as expected.  

\subsection{Structure factors}
\begin{figure}[ht]
\centering
\includegraphics[width=3in]{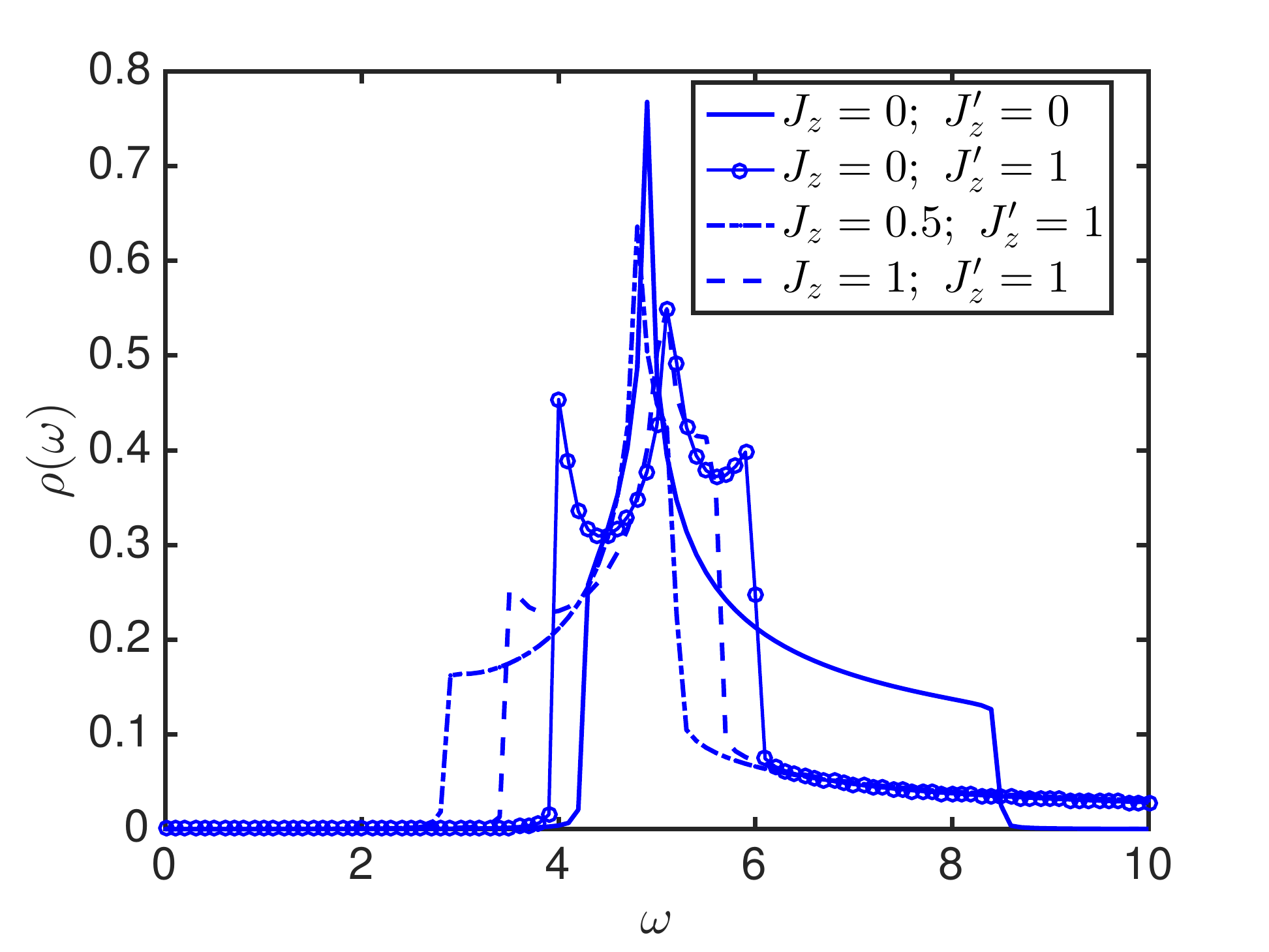}
\caption{Color online. The density of states $\rho(\omega)$ vs. $\omega$ at half-filling, $h_{x,z}=0~(\rho=0.5)$; $J_{\pm\pm}=1$.}
\label{DOS}
\end{figure}
\begin{figure}[ht]
\centering
\includegraphics[width=3in]{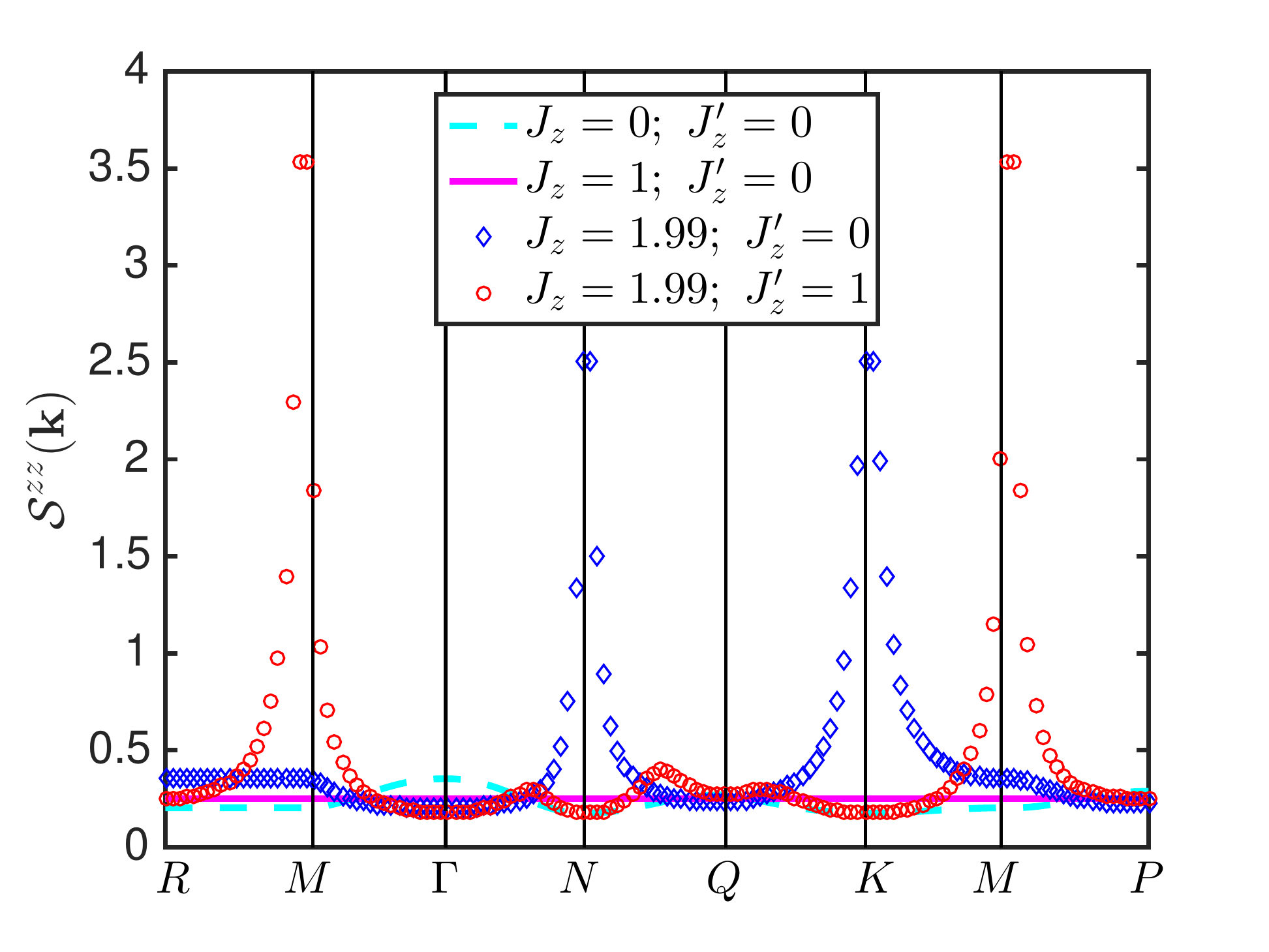}
\caption{Color online. The static dynamical structure factor, $\mathcal{S}^{zz}(\bo)$, along the Brillouin zone paths in Fig.~\eqref{K_spec1} at $h_{x,z}=0~(\rho=0.5)$;  $J_{\pm\pm}=1$ and $S=1/2$.}
\label{Tsf}
\end{figure} 
\begin{figure}[ht]
\centering
\includegraphics[width=3in]{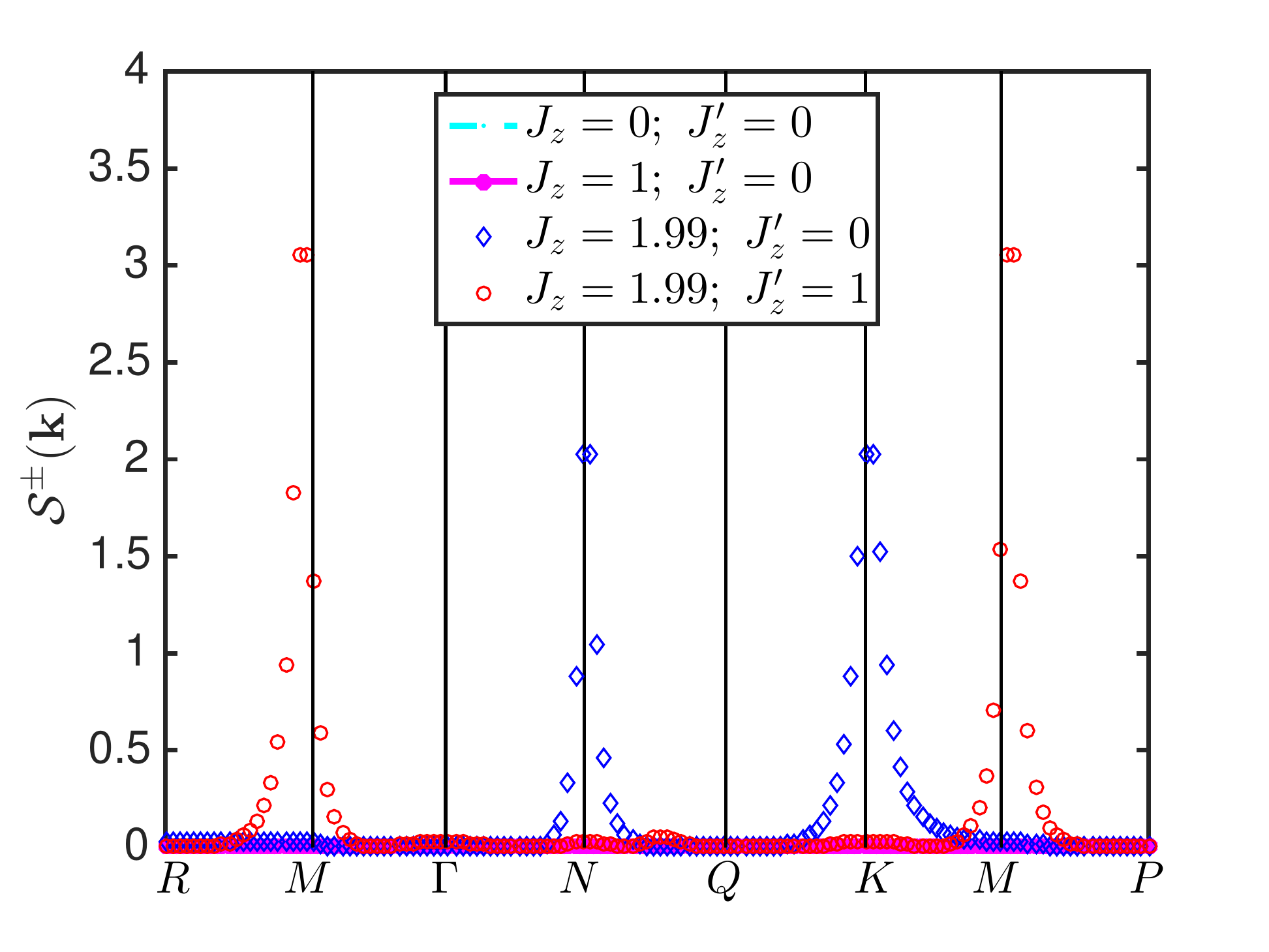}
\caption{Color online. The static dynamical structure factor,  $\mathcal{S}^{\pm}(\bo)$, along the Brillouin zone paths in Fig.~\eqref{K_spec1} at $h_{x,z}=0~(\rho=0.5)$;  $J_{\pm\pm}=1$ and $S=1/2$.}
\label{Tsf2}
\end{figure} 
\begin{figure}[ht]
\centering
\includegraphics[width=3in]{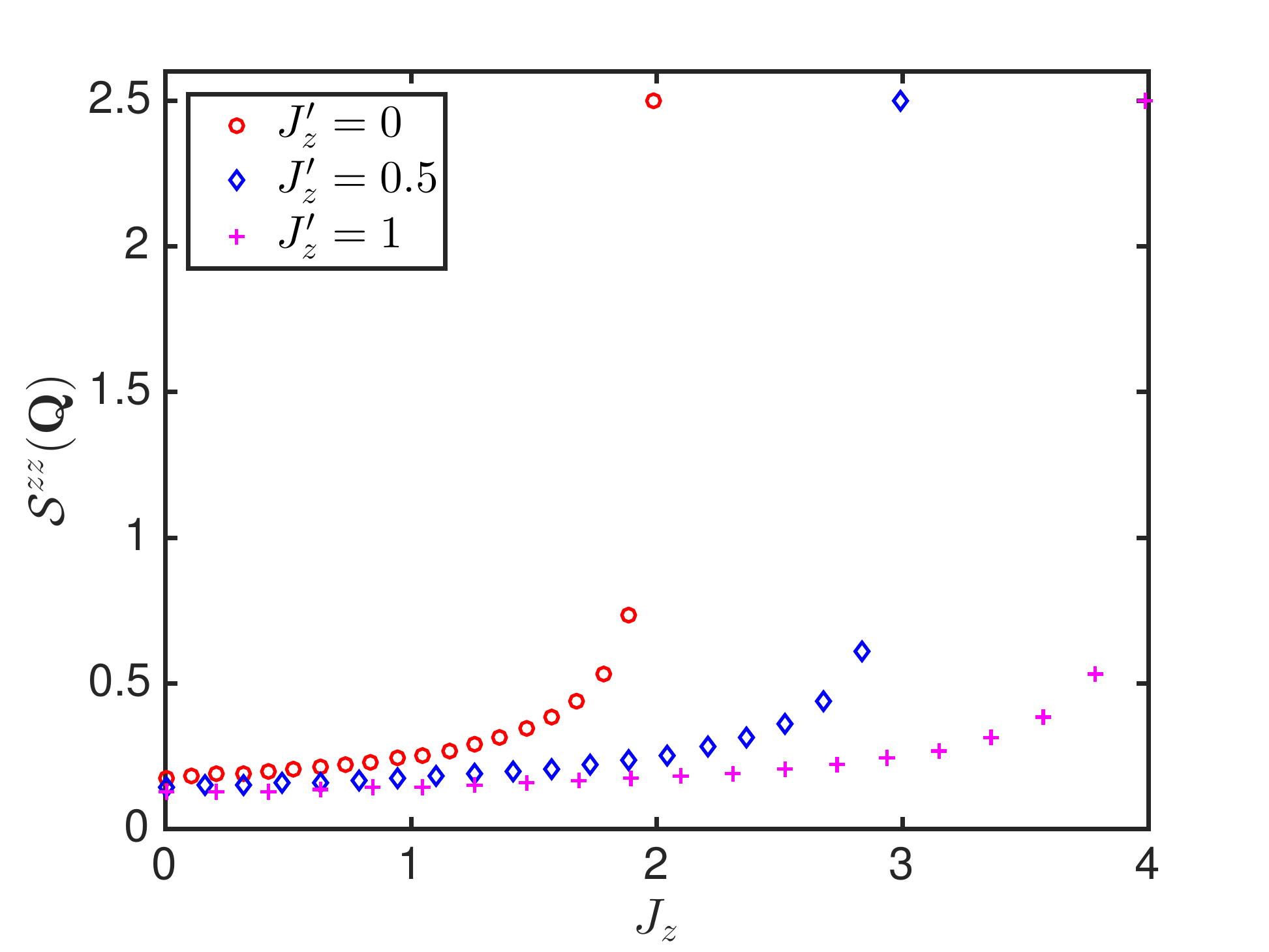}
\caption{Color online. The static structure factor,  $\mathcal{S}^{zz}(\bold{Q})$ as a function of $J_z$ at $J_{\pm\pm}=1$.}
\label{Sq}
\end{figure}

Due to the gapped nature of the $Z_2$-invariant model, all the thermodynamic quantities behave nicely without any divergent contributions. The density of states for this model is depicted in Fig.~\eqref{DOS} for several values of the anisotropies. There are some striking  features in the density of states of this model. We observe several spikes (van Hove singularities) depending on the anisotropies, which stem from the $\bo_\Gamma=0$ mode and the flat mode along $RM$. In the case of XY model, the major contribution to the  spike comes from the  $\bo_\Gamma=0$ mode and the discontinuity is a result of the lowest energy states at the roton minima  at the corners of the Brillouin zone.

Let us turn to the calculation of the dynamical structure factors, which is given by 
\begin{align}
\mathcal{S}^{\beta\gamma}(\bo,\omega)= \frac{1}{2\pi}\int_{-\infty}^{\infty}dt ~e^{i\omega t}\la S^\beta _{\bo}(t)S^\gamma_{-\bo}(0)\ra,
\end{align}
where $
S^\beta_{\bo}= \frac{1}{\sqrt N}\sum_l e^{-i\bo l}S_{l}^\beta
$ is the Fourier transform of the operators, and $\beta,\gamma=(x,y,z)$ label the components of the spins.  At zero temperature, the structure factors are obtained from the Green's function: \begin{align}
\mathcal{S}^{\beta\gamma}(\bo,\omega)=-\frac{1}{\pi}\text{Im}~\mathcal{G}^{\beta\gamma}(\bo,\omega),
\label{sfa}
\end{align}
with $\mathcal{G}^{\beta\gamma}(\bo,\omega)=-i\la \mathcal{T}  S^\beta_{\bo}(t)S^\gamma_{-\bo}(0)\ra$ being the time-ordered retarded Green's function. At half-filling or zero magnetic fields, $\theta=\pi/2$, Eq.~\eqref{trans} gives $S_l^x\to S_l^{\prime z}$ and  $S_l^z\to -S_l^{\prime x}$.  The quantization axis in this case is along the $x$-axis and the off-diagonal terms are $S_{l}^\pm = S_l^z\pm iS_l^y$. Using linear spin wave theory to order $S$ we find the two static structure factors at half-filling:  $\mathcal{S}^{zz}(\bo)= S(u_\bo -v_\bo)^2/2$ and $\mathcal{S}^{\pm}(\bo)= 2S v_\bo^2$. The static structure factors  $\mathcal{S}^{zz}(\bo)$ and $\mathcal{S}^{\pm}(\bo)$  are shown in Figs.~\eqref{Tsf}  and \eqref{Tsf2} respectively. In contrast to the $U(1)$-invariant model,  the  static  structure factors show no discontinuity in the entire Brillouin zone.  Our calculation shows that  $\mathcal{S}^{zz}(\bo)$ develops sharp step-like peaks at the minima of the energy spectra and it is completely flat at the rotationally symmetric point $J_{\pm\pm}=J_z=1$, $J_z^\prime=0$. The off-diagonal term  $\mathcal{S}^{\pm}(\bo)$   exhibits a similar behaviour with flat modes  except at the peaks and vanishes completely at the rotationally symmetric point, $J_{\pm\pm}=J_z=1$, $J_z^\prime=0$, as expected.  At $\bold{Q}=(\pm 4\pi/3,0)$ we find \bea \mathcal{S}^{zz}(\bold{Q})= \frac{S}{2}\sqrt{\frac{J_{\pm\pm}}{2(J_{\pm\pm}+J_z^\prime)-J_z}}.\eea
Figure \eqref{Sq} shows the plot of $\mathcal{S}^{zz}(\bold{Q})$ as a function of $J_z$.
%


\section{Conclusion} 
\label{sec9}
In this paper, we have uncovered the classical phase diagram   of the {\it quantum triangular ice}, and studied the effects of quantum fluctuations using spin wave theory.  We showed that, although the phase diagram is similar to the U(1)-invariant model, the interpretation of the phases are different due to the presence of $Z_2$ symmetry.  We presented the ground state energy at zero magnetic field and studied the trends of the order parameters at zero and nonzero magnetic fields. In the unfrustrated regime, we  explore the ground state thermodynamic properties of the easy-axis ferromagnetic phase. We observed some  features which are different from the U(1)-invariant XXZ model. In particular, divergent and discontinuous quantities  in the  U(1)-invariant XXZ model are finite in the $Z_2$-invariant  XXZ model, hence all the points in the Brillouin zone contribute to the thermodynamic quantities.   Interestingly, we found that linear spin wave theory provides an accurate picture of this system with the spin-deviation operator $\la n_l\ra<0.025$ for all parameter regimes considered. The particle density and the condensate fraction at half-filling give reasonable estimates at the level of our spin wave theory.  Also, the dynamical structure factors and the density of states exhibit interesting peaks at the minima of the energy spectra. 

In the frustrated regime, we have uncovered two phases --- a { \it ferrosolid} phase and a magnetized lobe. It would be interesting to numerically explore the {\it ferrosolid} state and corroborate the results of spin wave theory presented here. Experimentally, it might be interesting to study the nature of Eq.~\eqref{k1} in optical lattices  with ultra cold atoms.\cite{mar,duan} Quantum Monte Carlo (QMC) have uncovered a putative quantum spin liquid (QSL) phase in the magnetized lobes on the kagome lattice,\cite{juan} it would be interesting to numerically  explore  the nature of the magnetized lobes on the triangular lattice  and see if there are some interesting properties that could emerge from these lobes. The features uncovered in this paper might be useful for experimental purposes in gapped physical systems that could be modeled with the $Z_2$-invariant model.

%
%

\section*{ ACKNOWLEDGEMENTS} The author would like to thank Juan Carrasquilla for invaluable discussions. He also coined the name {\it ferrosolid}. I am also grateful to Anton Burkov and Roger  Melko for introducing me to this model. The author would also like to thank African Institute for Mathematical Sciences (AIMS), where  this work was completed. Research at Perimeter Institute is supported by the Government of Canada through Industry Canada and by the Province of Ontario through the Ministry of Research
and Innovation.

\appendix
\section{Spin Wave Theory}

 Spin wave theory  of Eq.~\eqref{k1} follows the standard approach.\cite{kle1, pw, jon, joli} Since the mean field theory is coplanar, we have $\phi_{\alpha\beta} =0$. However, in rotationally invariant systems, rotation about the $x$ or $y$ axis invariably yields the same results. In the  present model, there is no conserved quantity, thus rotation about the $x$-axis gives a different spin configuration from rotation about the $y$-axis. Hence, one has to rotate about the axis in which the presumed classical (mean field) energy is recovered as the order $S^2$ in the spin wave expansion. In our case, we rotate about the $y$-axis \begin{align}
&S_{l\alpha}^x=S_{l\alpha}^{\prime x}\cos\theta_\alpha +  S_{l\alpha}^{\prime z}\sin\theta_\alpha,\label{trans}\nonumber\\&
S_{l\alpha}^y=S_{l\alpha}^{\prime y},\\&\nonumber
S_{l\alpha}^z=- S_{l\alpha}^{\prime x}\sin\theta_\alpha + S_{l\alpha}^{\prime z}\cos\theta_\alpha.
\end{align} Next, we introduce a three-sublattice Holstein Primakoff transform.
 \begin{align}
 &S_{l\alpha}^{\prime z}= S-b_{l\alpha}^\dagger b_{ l\alpha}, \label{HP}\nonumber\\&
 S_{l\alpha}^{\prime y}= i\sqrt{\frac{S}{2}}\lb b_{l\alpha}^\dagger -b_{l\alpha}\rb,
 \\&\nonumber
 S_{l\alpha}^{\prime x}= \sqrt{\frac{S}{2}}\lb b_{l\alpha}^\dagger +b_{l\alpha}\rb.
 \end{align}
 Since the average spin-deviation operator $\la n_{l\alpha}\ra=\la b_{l\alpha}^\dagger b_{l\alpha}\ra$ is small, the linearized Holstein Primakoff transformation  is guaranteed.  Taking the magnetic fields, $h$, to be of order $S$ and keeping only the quadratic terms, after Fourier transform we obtain
\begin{align}
&H=\mathcal{E}_c+ S\sum_{\bo\alpha\beta}\lb \mathcal{M}_{\alpha\beta}^0\delta_{\alpha\beta} +\mathcal{M}_{\alpha\beta}^-\rb \lb b_{\bo \alpha}^\dagger b_{\bo \beta}+b_{-\bo \alpha}^\dagger b_{-\bo \beta}\rb\label{main}\\&\nonumber +\mathcal{M}_{\alpha\beta}^+ \lb b_{\bo \alpha}^\dagger b_{-\bo \beta}^\dagger +b_{-\bo \alpha} b_{\bo \beta}\rb,
\end{align}
where the coefficients are given by
\bea
\boldsymbol{\mathcal{M}}^0=\text{diag}\lb \xi_{AA},\xi_{BB},\xi_{CC}\rb,
\eea
\begin{align}
& \boldsymbol{\mathcal{M}}^\pm= \begin{pmatrix}
0& \lambda_{AB}^\pm\gamma_{AB}&\lambda_{CA}^\pm\gamma_{CA}^* \\
\lambda_{AB}^\pm\gamma_{AB}^*& 0& \lambda_{BC}^\pm\gamma_{BC}\\
 \lambda_{CA}^\pm\gamma_{CA}& \lambda_{BC}^\pm\gamma_{BC}^* & 0\\  
 \end{pmatrix},
\end{align} 
\bea \gamma_{AB}=\gamma_{CA}=\gamma_{BC} = \frac{1}{2}\sum_{j=1}^3 e^{ik_j};\quad k_j= \bold{k}\cdot \bold{e}_j,\eea

\begin{align}
&\xi_{\alpha\alpha}=\frac{h_\alpha-\sum_\beta z_{\alpha\beta}\lambda_{\alpha,\beta}^{1}}{2},\\&\lambda_{\alpha,\beta}^{\pm}=\frac{ \lambda_{\alpha,\beta}^{1x}\pm \lambda_{\alpha,\beta}^{1y}}{2},
\end{align}
where
\begin{align}
 &\lambda_{\alpha,\beta}^{1}=J_z\cos\theta_\alpha\cos\theta_\beta-J_{\pm\pm}\sin\theta_\alpha\sin\theta_\beta,\nonumber\\&  \lambda_{\alpha,\beta}^{1x}=J_z\sin\theta_\alpha\sin\theta_\beta-J_{\pm\pm}\cos\theta_\alpha\cos\theta_\beta,\nonumber\\& 
  \lambda_{\alpha,\beta}^{1y}=- J_{\pm\pm}; \quad h_\alpha= h\cos\theta_\alpha.
\end{align}
 The Hamiltonian can be written in terms of the Nambu operators $\Psi^\dg_\bo= (\psi^\dg_\bo, \thinspace \psi_{-\bo} )$  and $\psi^\dg_\bo=(b_{\bo A}^{\dg}\thinspace b_{\bo B}^{\dg}\thinspace b_{\bo C}^{\dg})$
\begin{align}
&H=\mathcal{E}_c+S\sum_{\bo} \Psi^\dg_\bo \mathcal{H}(\bo)\Psi_\bo-\sum_{\bo,\alpha}\xi_{\alpha\alpha},
\label{hp}
\end{align}

\begin{align}
\mathcal{H} = \mathbf{I}_{2\times 2}\otimes\lb \boldsymbol{\mathcal{M}}_0 +\boldsymbol{\mathcal{M}}_- \rb + \boldsymbol{\tau}_x\otimes  \boldsymbol{\mathcal{M}}_{+},
\end{align} 
where $\mathbf{I}_{2\times 2}$ is a $2\times 2$ identity matrix and $\boldsymbol{\tau}_x$ is a pseudo Pauli matrix.  
 The Hamiltonian can be diagonalized by the generalized Bogoliubov transformation \cite{jean,jean1, joli} 
 \bea\Psi(\bo)= \mathcal{U}Q(\bo),\eea 
 where $Q^\dg_\bo= (\mathcal{Q}_\bo^\dg,\thinspace \mathcal{Q}_{-\bo})$ with $ \mathcal{Q}_\bo^\dg=(\beta_{\bo A}^{\dg}\thinspace \beta_{\bo B}^{\dg}\thinspace \beta_{\bo C}^{\dg})$ being the quasiparticle operators. $\mathcal{U}$ is a $6\times 6$ matrix, which can be written as a $2\times 2$ matrix
 \begin{align}
& \mathcal{U}= \begin{pmatrix}
  u& -v^* \\
-v&u^*\\  
 \end{pmatrix},
\end{align} 
where $u,~v$ are  $3\times 3$ matrices that satisfy $|u|^2-|v|^2=1$. 
 If the Hamiltonian in Eq.~\eqref{hp} is to be diagonalized, then $\mathcal{U}$ has to satisfy the relation
\begin{align}
&\mathcal{U}^\dg \mathcal{H} \mathcal{U}= \epsilon(\bo); \quad \mathcal{U}^\dg \zeta \mathcal{U}= \zeta,
\label{eig}
\end{align}
where
$\zeta =
\text{diag}(
 \mathbf{I}_{3\times 3}, -\mathbf{I}_{3\times 3} )$, and $\epsilon(\bo)$ is the diagonal matrix of the bosonic eigenvalues. It follows from Eq.~\eqref{eig}, that we need to numerically diagonalize a non-Hermitian matrix $\zeta \mathcal{H}$, whose eigenvalues are given by $\zeta \epsilon(\bo)=[\epsilon_\alpha(\bo), -\epsilon_\alpha(\bo)]$ and the columns of $\mathcal U$ are the corresponding eigenvectors. The diagonalized form is given by 
\begin{align}
H=\mathcal{E}_g +S\sum_{\bo,\alpha}  \epsilon_\alpha(\bo)\lb\beta_{\bo\alpha}^\dg\beta_{\bo\alpha} +\beta_{-\bo\alpha}^\dg\beta_{-\bo\alpha}\rb,
\end{align}
where
$\epsilon_\alpha(\bo)$ are obtained numerically and the correction to the classical energy is given by
\begin{align}
\Delta\mathcal{E}=\mathcal{E}_g-\mathcal{E}_c=S\sum_{\bo,\alpha}[\epsilon_\alpha(\bo)-\xi_{\alpha\alpha}].
\label{gr}
\end{align}


\end{document}